\documentclass[lettersize,journal]{IEEEtran}

\usepackage{cite}
\usepackage{amsmath,amssymb,amsfonts}
\usepackage{algorithm}
\usepackage{textcomp}
\usepackage{graphicx}

\usepackage{acronym} 
\acrodef{USPA}[USPA]{uniform square planar array} 

\usepackage[colorlinks,linkcolor=blue,anchorcolor=blue,citecolor=blue]{hyperref} 
\usepackage{orcidlink}

\usepackage{color}
\usepackage{xcolor}
\usepackage{caption}
\usepackage[noend]{algpseudocode}
\usepackage[caption=false,font=footnotesize,labelfont=rm,textfont=rm]{subfig}
\def\BibTeX{{\rm B\kern-.05em{\sc i\kern-.025em b}\kern-.08em
    T\kern-.1667em\lower.7ex\hbox{E}\kern-.125emX}}

\makeatletter
\renewcommand{\maketag@@@}[1]{\hbox{\m@th\normalsize\normalfont#1}}
\makeatother

\begin{document}

\title{AREE-Based Decoupled Design of Hybrid Beamformers in mmWave XL-MIMO Systems}

\author{Jiazhe Li$^{\orcidlink{0000-0002-1042-9981}}$,~\IEEEmembership{Graduate Student Member,~IEEE}, Nicolò Decarli$^{\orcidlink{0000-0002-0359-8808}}$,~\IEEEmembership{Member,~IEEE}, Francesco Guidi$^{\orcidlink{0000-0002-1773-8541}}$,~\IEEEmembership{Member,~IEEE}, Heng Dong$^{\orcidlink{0000-0001-8961-0531}}$, Anna Guerra$^{\orcidlink{0000-0001-5214-1444}}$,~\IEEEmembership{Member,~IEEE}, Alessandro Bazzi$^{\orcidlink{0000-0003-3500-1997}}$,~\IEEEmembership{Senior Member,~IEEE}, and Zhuoming Li$^{\orcidlink{0000-0001-5172-6256}}$



\thanks{

This work was supported by the China Scholarship Council under Grant No.202406120095.

\textit{(Corresponding author: Zhuoming Li.)}
  
Jiazhe Li, Heng Dong and Zhuoming Li are with the Department of Electronics and Information Engineering, Harbin Institute of Technology, Harbin 150001, China.

Nicolò Decarli and Francesco Guidi are with the National Research Council, Institute of Electronics, Computer and Telecommunication Engineering (CNR-IEIIT) and WiLab-CNIT, 40136 Bologna, Italy.

Alessandro Bazzi and Anna Guerra are with the University of Bologna and WiLab-CNIT, 40136 Bologna, Italy.
}

}

\markboth{}%
{Shell \MakeLowercase{\textit{et al.}}: A Sample Article Using IEEEtran.cls for IEEE Journals}


\maketitle
\begin{abstract}
Hybrid beamforming has been widely employed in mmWave communications such as vehicular-to-everything (V2X) scenarios, as a compromise between hardware complexity and spectral efficiency. However, the inherent coupling between analog and digital precoders in hybrid array architecture significantly limits the computational and spectral efficiency of existing algorithms. To address this issue, we propose an alternating residual error elimination (AREE) algorithm, which decomposes the hybrid beamforming problem into two low-dimensional subproblems, each exhibiting a favorable matrix structure that enables effective decoupling of analog and digital precoders from the matrix product formulation. These subproblems iteratively eliminate each other's residual errors, driving the original problem toward the optimal hybrid beamforming performance. The proposed initialization ensures rapid convergence, while a low-complexity geometric channel SVD algorithm is developed by transforming the high-dimensional sparse channel into a low-dimensional equivalent, thereby simplifying the derivation of subproblems. Simulation results demonstrate that the AREE algorithm effectively decouples analog and digital precoders with low complexity, achieves fast convergence, and offers higher spectral efficiency than existing beamforming methods.
\end{abstract}


\section{Introduction}
\IEEEPARstart{A}{s} wireless communication evolves from sub-6 GHz to mmWave bands, the shrinking size of antenna elements to millimeter scale facilitates the integration of extremely large-scale MIMO (XL-MIMO) systems \cite{cui2022channel,li2024fast}. MmWave XL-MIMO enhances angular resolution and spectral efficiency, thereby advancing integrated sensing and communications (ISAC) particularly in vehicular-to-everything (V2X) scenarios \cite{liu2022integrated,bartoletti2024integration,decarli2023v2x}. However, mmWave signals experience more severe attenuation than lower frequencies \cite{akdeniz2014millimeter,heath2016overview}, resulting in limited reflected paths and channel spatial sparsity \cite{alkhateeb2014channel,lee2016channel,rodriguez2018frequency}. To address these challenges, mmWave communications employ beamforming to compensate for signal attenuation and adapt to channel sparsity via highly directional beams \cite{elbir2023twenty,li2023hybrid}. However, XL-MIMO typically employs hybrid array architectures \cite{el2014spatially,ahmed2018survey}, in which all antennas share a limited number of radio frequency (RF) chains to reduce power consumption and hardware complexity. This configuration imposes a ``constant modulus constraint" \cite{el2014spatially,ahmed2018survey,li2023hybrid}, allowing only phase control for each antenna connected to an RF chain. While hybrid beamforming is less spectrally efficient than fully-digital beamforming (which achieves optimal performance through fully-digital array architectures \cite{palomar2003joint,viamimo}), it remains crucial for practical implementation. Consequently, to fully exploit advantages of hybrid array architectures, considerable research has focused on hybrid beamforming algorithms that aim to approximate the performance of fully-digital architectures.

However, the hybrid array architecture consists of coupled analog and digital precoders \cite{el2014spatially,ahmed2018survey}, whose matrix product formulation renders hybrid beamforming mathematically intractable. Although decoupling these precoders from the matrix product formulation introduces decoupling errors that compromise spectral efficiency \cite{wang2022joint,el2014spatially,lee2014hybrid,chen2016compressive,tsai2018sub,zilli2021constrained,sohrabi2016hybrid,yu2016alternating,sohrabi2017hybrid,ioushua2019family}, maintaining coupled precoders leads to prohibitive computational complexity for hybrid beamforming \cite{yu2016alternating,lin2019hybrid,hui2023hybrid,yuan2023alternating,zhu2023max,yu2022regularized,ci2025hybrid,liu2025dynamic}. The effectively decoupled analog and digital precoders would simultaneously improve both computational efficiency and spectral efficiency, yet no existing decoupling algorithm achieves this without introducing decoupling errors. This unresolved challenge constitutes a critical research gap in the current literature.

In recent years, numerous studies have focused on hybrid beamforming with multiple antennas at both the transmitter and receiver \cite{wang2022joint,el2014spatially,lee2014hybrid,chen2016compressive,lin2019hybrid,hui2023hybrid,yuan2023alternating,zhu2023max,sohrabi2016hybrid,yu2016alternating,sohrabi2017hybrid,ioushua2019family}. The design process typically involves two sequential steps: first design the transmitter precoders, followed by the design of receiver combiners. Since the algorithms for transmitters are fully applicable to receivers \cite{el2014spatially,yu2016alternating}, most studies focus on transmitter precoder design, which can be broadly categorized into non-iterative and iterative algorithms. Non-iterative algorithms \cite{wang2022joint,el2014spatially,lee2014hybrid,chen2016compressive,tsai2018sub,zilli2021constrained} typically employ simple decoupling methods such as orthogonal projection \cite{el2014spatially}, with their fixed number of computational steps ensuring low complexity. Among these, Orthogonal Matching Pursuit (OMP)-based algorithms derived from compressed sensing theory are widely studied \cite{el2014spatially,lee2014hybrid,chen2016compressive}. These algorithms project the optimal fully-digital solution onto the analog precoder and then select beam components with the largest projections to reconstruct both analog and digital precoders \cite{el2014spatially}. Computational efficiency is further improved in \cite{lee2014hybrid,chen2016compressive} using methods such as Schur-Banachiewicz blockwise inversion. However, these algorithms \cite{el2014spatially,lee2014hybrid,chen2016compressive} suffer from performance degradation because the orthogonal projection limits the number of selectable beams for precoder reconstruction, introducing significant decoupling errors. Other algorithms such as those in \cite{tsai2018sub,zilli2021constrained,wang2022joint} optimize the analog and digital precoders separately without feedback, inevitably introducing decoupling errors that degrade overall performance. Although computationally efficient, non-iterative algorithms fail to effectively suppress decoupling errors, ultimately compromising spectral efficiency.   

Iterative algorithms \cite{lin2019hybrid,hui2023hybrid,yuan2023alternating,zhu2023max,sohrabi2016hybrid,yu2016alternating,sohrabi2017hybrid,ioushua2019family} perform better in spectral efficiency because they decouple precoders more effectively than their non-iterative counterparts, which can be classified into two categories. The first category decouples analog and digital precoders from their matrix product formulation before iteratively optimizing them to approximate the optimal solution \cite{sohrabi2016hybrid,yu2016alternating,sohrabi2017hybrid,ioushua2019family}. However, the row rank-deficient nature of digital precoders complicates the decoupling process, prompting various solutions. For instance, the PE-AltMin algorithm \cite{yu2016alternating} approximates the right inverse of the digital precoder using its conjugate transpose, while \cite{sohrabi2016hybrid,sohrabi2017hybrid} models the digital precoder as a block matrix combining a unitary matrix and a zero matrix to simplify computations. A more flexible approach in \cite{ioushua2019family} assumes an identity matrix digital precoder for easier decoupling, introducing an intermediate variable to enhance optimization flexibility. Although these algorithms \cite{sohrabi2016hybrid,yu2016alternating,sohrabi2017hybrid,ioushua2019family} decouple precoders efficiently, they introduce significant decoupling errors that hinder effective precoder optimization, resulting in performance gaps relative to the optimal solution. The second category \cite{yu2016alternating,lin2019hybrid,hui2023hybrid,yuan2023alternating,zhu2023max,yu2022regularized,ci2025hybrid,liu2025dynamic} avoids decoupling from the matrix product formulation entirely, instead transforming the matrix product into a high-dimensional vector and applying advanced mathematical optimization methods to directly solve the coupled precoders. Examples include the MO-AltMin algorithm \cite{yu2016alternating,lin2019hybrid,hui2023hybrid} employing Riemannian manifold and gradient descent methods, and the AO-MM algorithm \cite{yuan2023alternating,zhu2023max,yu2022regularized,ci2025hybrid,liu2025dynamic} applying max-min optimization theory. While these algorithms typically achieve satisfactory spectral efficiency by avoiding decoupling errors through direct optimization of coupled precoders, they incur substantial computational complexity that limits their practical applicability.

Despite extensive research on hybrid beamforming, effectively decoupling analog and digital precoders in hybrid array architectures remains a critical challenge. Existing algorithms \cite{wang2022joint,el2014spatially,lee2014hybrid,chen2016compressive,tsai2018sub,zilli2021constrained,sohrabi2016hybrid,yu2016alternating,sohrabi2017hybrid,ioushua2019family} typically introduce significant decoupling errors that degrade performance, while alternative approaches \cite{yu2016alternating,lin2019hybrid,hui2023hybrid,yuan2023alternating,zhu2023max,yu2022regularized,ci2025hybrid,liu2025dynamic} that directly optimize coupled precoders result in prohibitive computational complexity. An effective decoupling algorithm would simultaneously improve both computational and spectral efficiency. These considerations motivated our exploration of novel decoupling algorithms for hybrid beamforming.

\begin{figure*}[t]
  \centering
  \subfloat[Fully-connected hybrid array architecture for hybrid beamforming]
  {\label{fig1a}\includegraphics[width=0.7\textwidth]{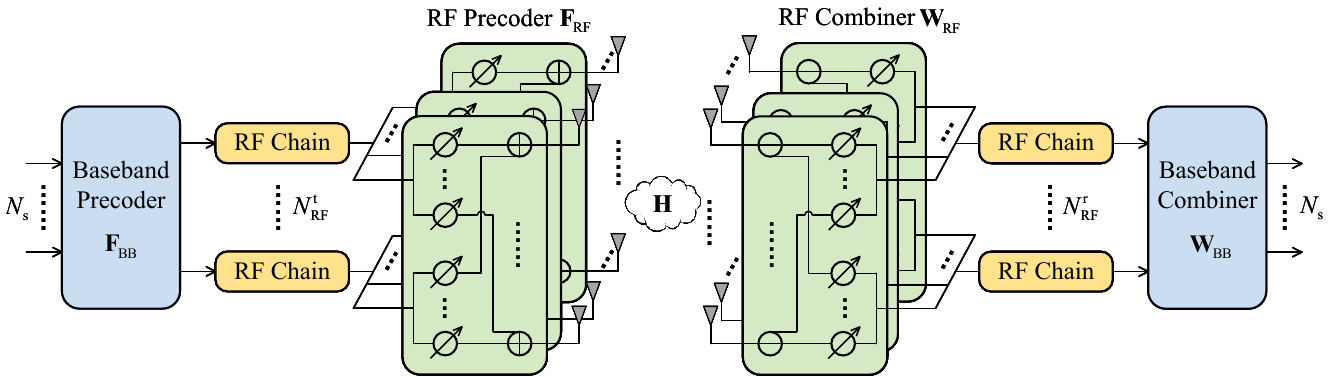}}
  \hspace{1cm} 
  \subfloat[Uniform square planar array]
  {\label{fig1b}\includegraphics[width=0.2\textwidth]{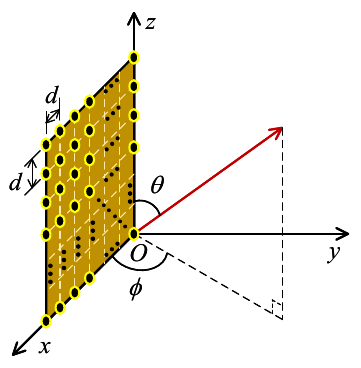}}
  \caption{Hybrid beamforming for mmWave XL-MIMO communication systems with USPA in Cartesian coordinate.}
  \label{fig1}
\end{figure*}

In this paper, we propose a novel alternating residual error elimination (AREE) algorithm that decouples analog and digital precoders effectively in hybrid array architectures, introducing negligible decoupling errors and thus achieving improved spectral efficiency while maintaining low complexity of hybrid beamforming. The main contributions of this paper are summarized as follows:

\begin{itemize}
\item First, we introduce the novel AREE algorithm to effectively decouple analog and digital precoders in hybrid array architectures. It decomposes the hybrid beamforming problem into two low-dimensional subproblems, each exhibiting a favorable matrix structure for precoder decoupling with negligible decoupling errors. These subproblems alternately eliminate each other's residual errors, driving the original problem toward the optimal hybrid beamforming performance. The proposed AREE algorithm achieves higher spectral efficiency than existing algorithms while maintaining low complexity.

\item Second, to efficiently derive objective functions of the AREE algorithm in hybrid beamforming, we propose a low-complexity geometric channel SVD (GC-SVD) method that transforms high-dimensional sparse channel into a low-dimensional representation by exploiting spatial sparsity of mmWave XL-MIMO channel. It significantly reduces complexity of channel decomposition while simplifying initialization of the AREE algorithm.

\item Finally, to accelerate convergence of the AREE algorithm, we propose two initialization methods named PE-OMP and PE-SMD. The PE-OMP algorithm leverages both the strongest beam components and the remaining components for precoder reconstruction, achieving superior performance compared to existing OMP-based algorithms while maintaining lower complexity. The PE-SMD algorithm further simplifies PE-OMP with only marginal performance degradation, providing effective initial values for the AREE algorithm.
\end{itemize}

The rest of this paper is organized as follows. Sec.~II presents system model and problem formulation. The AREE algorithm, low complexity SVD and its initialization are proposed in Sec.~III, Sec.~IV and Sec.~V, respectively. In Sec.~VI and Sec.~VII, computational complexity and simulation results are provided. Finally, we conclude this paper in Sec.~VIII.

\textit{Notations}: The conjugate, transpose and conjugate transpose of $\mathbf{A}$ are denoted as ${{\mathbf{A}}^{*}}$, ${{\mathbf{A}}^{T}}$, and ${{\mathbf{A}}^{H}}$, respectively. ${{\left[ \mathbf{A} \right]}_{i:j,\ :}}$ (or ${{\left[ \mathbf{A} \right]}_{:,\ i:j}}$) represents a matrix consisting of the elements of $\mathbf{A}$ from the $i$th row (or column) to the $j$th row (or column), while ${{\left[ \mathbf{A} \right]}_{i,j}}$ is the entry on the $i$th row and $j$th column of $\mathbf{A}$. $\operatorname{Tr}\left( \mathbf{A} \right)$, $\left| \mathbf{A} \right|$, and ${{\left\| \mathbf{A} \right\|}_{F}}$ represent the trace, determinant and Frobenius norm of $\mathbf{A}$. ${{\mathbf{A}}^{-1}}$ is the inverse of $\mathbf{A}$, and ${{\mathbf{A}}^{\dagger }}$ denotes the Moore-Penrose pseudo inverse. $\operatorname{rank}\left( \mathbf{A} \right)$ calculates the rank of $\mathbf{A}$, while $\arg \left( \mathbf{A} \right)$ is a matrix composed of the phase of $\mathbf{A}$. $\operatorname{diag}\left( \mathbf{a} \right)$ transforms the vector $\mathbf{a}$ into a diagonal matrix. ${{\mathbf{I}}_{N}}$ is a $N\times N$ identity matrix. The semi-unitary matrix $\mathbf{A}\in {{\mathbb{C}}^{M\times N}}$ is defined as satisfying ${{\mathbf{A}}^{H}}\mathbf{A}={{\mathbf{I}}_{N}}$ when $M\ge N$, or ${\mathbf{A}}{\mathbf{A}}^{H}={{\mathbf{I}}_{M}}$ when $N\ge M$. $Laplace\left( \phi ,{\sigma }/{\sqrt{2}}\; \right)$ denotes the Laplace distribution with a mean of $\phi $ and a standard deviation of $\sigma $, respectively.

\section{System Model}
\subsection{System Model}
As illustrated in Fig. \ref{fig1}, we study a single-user mmWave XL-MIMO downlink narrowband communication system\footnote{The decoupled design of hybrid precoders proposed in this paper is also applicable to wideband scenarios, such as OFDM and multi-user systems \cite{zilli2021constrained}.}. The base station (BS) and mobile station (MS) both employ a fully-connected hybrid array architecture \cite{el2014spatially}. The BS is equipped with ${{N}_{\mathrm{t}}}$ antennas and $N_{\mathrm{RF}}^{\mathrm{t}}$ RF chains, while the MS has ${{N}_{\mathrm{r}}}$ antennas and $N_{\mathrm{RF}}^{\mathrm{r}}$ RF chains, satisfying $N_{\mathrm{RF}}^{\mathrm{t}}<2{{N}_{\mathrm{s}}}$ and $N_{\mathrm{RF}}^{\mathrm{r}}<2{{N}_{\mathrm{s}}}$ where ${{N}_{\mathrm{s}}}$ represents the number of data streams being transmitted. It is assumed that ${{N}_{\mathrm{s}}}\le N_{\mathrm{RF}}^{\mathrm{t}}\ll {{N}_{\mathrm{t}}}$ and ${{N}_{\mathrm{s}}}\le N_{\mathrm{RF}}^{\mathrm{r}}<{{N}_{\mathrm{r}}}$ to ensure communication effectiveness.

Denote the data symbols transmitted from BS to MS as $\mathbf{s}\in {{\mathbb{C}}^{{{N}_{\mathrm{s}}}\times 1}}$, which satisfies $\mathbb{E}\left[ \mathbf{s}{{\mathbf{s}}^{H}} \right]=\left( {{{P}_{\mathrm{t}}}}/{{{N}_{\mathrm{s}}}}\; \right){{\mathbf{I}}_{{{N}_{\mathrm{s}}}}}$ and ${{P}_{\mathrm{t}}}$ represents the total transmission power. The baseband signal $\mathbf{s}$ is first processed by the baseband digital precoder ${{\mathbf{F}}_{\mathrm{BB}}}\in {{\mathbb{C}}^{N_{\mathrm{RF}}^{\mathrm{t}}\times {{N}_{\mathrm{s}}}}}$, followed with up-conversion by the RF analog precoder ${{\mathbf{F}}_{\mathrm{RF}}}\in {{\mathbb{C}}^{{{N}_{\mathrm{t}}}\times N_{\mathrm{RF}}^{\mathrm{t}}}}$ before finally being transmitted through the antenna array. Similarly, the received signal at MS is first down-converted to the baseband by RF analog combiner ${{\mathbf{W}}_{\mathrm{RF}}}\in {{\mathbb{C}}^{{{N}_{\mathrm{r}}}\times N_{\mathrm{RF}}^{\mathrm{r}}}}$, and then processed by the baseband digital combiner ${{\mathbf{W}}_{\mathrm{BB}}}\in {{\mathbb{C}}^{N_{\mathrm{RF}}^{\mathrm{r}}\times {{N}_{\mathrm{s}}}}}$ to get the final received signal. We consider a complex baseband model in this paper, and the received signal is expressed as
\begin{equation}
  \mathbf{y}=\mathbf{W}_{\mathrm{BB}}^{H}\mathbf{W}_{\mathrm{RF}}^{H}\mathbf{H}{{\mathbf{F}}_{\mathrm{RF}}}{{\mathbf{F}}_{\mathrm{BB}}}\mathbf{s}+\mathbf{W}_{\mathrm{BB}}^{H}\mathbf{W}_{\mathrm{RF}}^{H}\mathbf{n},
  \label{eq1}
\end{equation}
where $\mathbf{H}\in {{\mathbb{C}}^{{{N}_{\mathrm{r}}}\times {{N}_{\mathrm{t}}}}}$ and $\mathbf{n}\sim \mathcal{C}\mathcal{N}\left( \mathbf{0},\sigma _{\mathrm{n}}^{2}{{\mathbf{I}}_{{{N}_{\mathrm{r}}}}} \right)$ represent the mmWave channel and Gaussian white noise. Both analog precoder and combiner are implemented by analog phase shifters \cite{el2014spatially}, resulting in ``constant modulus constraint" as $\left| {{\left[ {{\mathbf{F}}_{\mathrm{RF}}} \right]}_{i,j}} \right|={1}/{\sqrt{{{N}_{\mathrm{t}}}}}\;$ and $\left| {{\left[ {{\mathbf{W}}_{\mathrm{RF}}} \right]}_{i,j}} \right|={1}/{\sqrt{{{N}_{\mathrm{r}}}}}\;$. Additionally, ${{\mathbf{F}}_{\mathrm{BB}}}$ is normalized to satisfy the total transmission power constraint of BS as $\left\| {{\mathbf{F}}_{\mathrm{RF}}}{{\mathbf{F}}_{\mathrm{BB}}} \right\|_{F}^{2}={{N}_{\mathrm{s}}}$. When Gaussian symbols are transmitted over the mmWave channel, the spectral efficiency achieved at MS is given by \cite{el2014spatially,yu2016alternating}
\begin{equation}
\begin{aligned}
  & R={{\log }_{2}}\left( \left| {{\mathbf{I}}_{{{N}_{\mathrm{s}}}}}+\frac{{{P}_{\mathrm{t}}}}{{{N}_{\mathrm{s}}}}\mathbf{R}_{\mathrm{n}}^{-1}\mathbf{W}_{\mathrm{BB}}^{H}\mathbf{W}_{\mathrm{RF}}^{H}\mathbf{H}{{\mathbf{F}}_{\mathrm{RF}}}{{\mathbf{F}}_{\mathrm{BB}}} \right. \right. \\ 
 & \left. \left. \quad \quad \quad \quad \quad \quad \quad \quad \quad \times \mathbf{F}_{\mathrm{BB}}^{H}\mathbf{F}_{\mathrm{RF}}^{H}{{\mathbf{H}}^{H}}{{\mathbf{W}}_{\mathrm{RF}}}{{\mathbf{W}}_{\mathrm{BB}}} \right| \right), \\ 
\end{aligned}
\label{eq2}
\end{equation}
where ${{\mathbf{R}}_{\mathrm{n}}}=\sigma _{\mathrm{n}}^{2}\mathbf{W}_{\mathrm{BB}}^{H}\mathbf{W}_{\mathrm{RF}}^{H}{{\mathbf{W}}_{\mathrm{RF}}}{{\mathbf{W}}_{\mathrm{BB}}}\in {{\mathbb{C}}^{{{N}_{\mathrm{s}}}\times {{N}_{\mathrm{s}}}}}$ represents the covariance matrix of the white noise after beam combining.

\subsection{MmWave XL-MIMO Sparse Channel Model}
Assume that both BS and MS employ \acp{USPA} with ${{N}_{\mathrm{t}}}=\sqrt{{{N}_{\mathrm{t}}}}\times \sqrt{{{N}_{\mathrm{t}}}}$ and ${{N}_{\mathrm{r}}}=\sqrt{{{N}_{\mathrm{r}}}}\times \sqrt{{{N}_{\mathrm{r}}}}$ antennas, respectively. The inter-element spacing of the USPA is $d={\lambda }/{2}\;$ where $\lambda $ represents the carrier wavelength, and then the downlink mmWave geometric channel can be expressed as \cite{el2014spatially,sohrabi2016hybrid,chen2016compressive,yu2016alternating}:
\begin{equation}
  \mathbf{H}=\sqrt{\frac{{{N}_{\mathrm{t}}}{{N}_{\mathrm{r}}}}{{{N}_{\mathrm{cl}}}{{N}_{\mathrm{ray}}}}}\sum\limits_{i=1}^{{{N}_{\mathrm{cl}}}}{\sum\limits_{l=1}^{{{N}_{\mathrm{ray}}}}{{{\alpha }_{i,l}}{{\mathbf{a}}_{\mathrm{r}}}\left( \phi _{i,l}^{\mathrm{r}},\theta _{i,l}^{\mathrm{r}} \right){{\mathbf{a}}_{\mathrm{t}}}{{\left( \phi _{i,l}^{\mathrm{t}},\theta _{i,l}^{\mathrm{t}} \right)}^{H}}}},
  \label{eq3}
\end{equation}
where $\mathbf{H}\in {{\mathbb{C}}^{{{N}_{\mathrm{r}}}\times {{N}_{\mathrm{t}}}}}$ satisfies $\mathbb{E}\left[ \left\| \mathbf{H} \right\|_{F}^{2} \right]={{N}_{\mathrm{t}}}{{N}_{\mathrm{r}}}$, and we assume $\operatorname{rank}\left( \mathbf{H} \right)\ge {{N}_{\mathrm{s}}}$ to ensure the feasibility of transmitting ${{N}_{\mathrm{s}}}$ data streams with satisfied communication quality. ${{N}_{\mathrm{cl}}}$ and ${{N}_{\mathrm{ray}}}$ denote the number of scattering clusters and propagation paths (rays) in each cluster, and ${{\alpha }_{i,l}}\sim \mathcal{C}\mathcal{N}\left( 0,1 \right)$ is the complex channel gain of the $l$th propagation path in the $i$th cluster. $\phi _{i,l}^{\mathrm{r}}$ ($\phi _{i,l}^{\mathrm{t}}$) and $\theta _{i,l}^{\mathrm{r}}$ ($\theta _{i,l}^{\mathrm{t}}$) denote the azimuth angle and elevation angle of arrival (departure), respectively. ${{\mathbf{a}}_{\mathrm{t}}}\left( \phi _{i,l}^{\mathrm{t}},\theta _{i,l}^{\mathrm{t}} \right)\in {{\mathbb{C}}^{{{N}_{\mathrm{t}}}\times 1}}$ and ${{\mathbf{a}}_{\mathrm{r}}}\left( \phi _{i,l}^{\mathrm{r}},\theta _{i,l}^{\mathrm{r}} \right)\in {{\mathbb{C}}^{{{N}_{\mathrm{r}}}\times 1}}$ are the normalized steering vectors of the transmitter and receiver. As shown in Fig. \ref{fig1}\subref{fig1b}, supposing that the USPA is located on the $x$-$z$ plane of the Cartesian coordinate \cite{qin2022partial}, the steering vector can be expressed as \cite{el2014spatially,yu2016alternating}: 
\begin{equation}
\begin{aligned}
  & \mathbf{a}\left( {{\phi }_{i,l}},{{\theta }_{i,l}} \right)\!=\!\frac{1}{\sqrt{N}}\!\left[ 1,\ldots ,{{e}^{\mathrm{j}\!\frac{2\pi d}{\lambda }\left( {{n}_{\mathrm{h}}}\!\cos \left( {{\phi }_{i,l}} \right)\sin \left( {{\theta }_{i,l}} \right)+{{n}_{\mathrm{v}}}\!\cos \left( {{\theta }_{i,l}} \right) \right)}}, \right. \\ 
 & {{\left. \quad \quad \quad \ \ \ \,\ldots ,{{e}^{\mathrm{j}\!\frac{2\pi d}{\lambda }\left( \left( \!\sqrt{\!N}-1 \right)\!\cos \left( {{\phi }_{i,l}} \right)\sin \left( {{\theta }_{i,l}} \right)+\left( \!\sqrt{\!N}-1 \!\right)\!\cos \left( {{\theta }_{i,l}} \right) \right)}} \right]}^{\!T}}, \\ 
\end{aligned}
\label{eq4}
\end{equation}
where we have used a unified form (ignoring labels $\mathrm{r}$ and $\mathrm{t}$). $0\le {{n}_{\mathrm{h}}}<\sqrt{N}$ and $0\le {{n}_{\mathrm{v}}}<\sqrt{N}$ represent the antenna indices along the horizontal and vertical directions of the USPA, respectively. Ignoring the subscripts $\mathrm{r}$ and $\mathrm{t}$, we model ${{\phi }_{i,l}}\sim Laplace\left( {{\phi }_{i}},{{{\sigma }_{\mathrm{\phi} }}}/{\sqrt{2}}\; \right)$ and ${{\theta }_{i,l}}\sim Laplace\left( {{\theta }_{i}},{{{\sigma }_{\mathrm{\theta} }}}/{\sqrt{2}}\; \right)$ as described in \cite{el2014spatially}, where ${{\phi }_{i}},{{\theta }_{i}}\!\!\sim\!\! \operatorname{U}\left( 0,\pi  \right)$ denote mean cluster angles of the $i$th cluster, and ${{\sigma }_{\mathrm{\phi} }}$, ${{\sigma }_{\mathrm{\theta} }}$ represent angle spreads.

\subsection{Problem Formulation}
To decouple the joint transmitter-receiver optimization, we begin by designing the transmitter precoding, noting that the following analysis is fully applicable to receiver combining as well \cite{yu2016alternating}. Instead of directly maximizing spectral efficiency, we optimize ${{\mathbf{F}}_{\mathrm{RF}}}{{\mathbf{F}}_{\mathrm{BB}}}$ to maximize the mutual information achieved by Gaussian signaling over the mmWave channel \cite{el2014spatially}:
\begin{equation}
  \mathcal{I}\left( {{\mathbf{F}}_{\mathrm{\!RF}}},{{\mathbf{F}}_{\mathrm{\!BB}}} \right)={{\log }_{2}}\!\left( \left| {{\mathbf{I}}_{{{N}_{\mathrm{r}}}}}\!+\!\frac{{{P}_{\mathrm{t}}}}{{{N}_{\mathrm{s}}}\sigma _{\mathrm{n}}^{2}}\mathbf{H}{{\mathbf{F}}_{\mathrm{\!RF}}}{{\mathbf{F}}_{\mathrm{\!BB}}}\mathbf{F}_{\mathrm{\!BB}}^{H}\mathbf{F}_{\mathrm{\!RF}}^{H}{{\mathbf{H}}^{H}} \right| \right),
\label{eq5}
\end{equation}
where the mutual information is maximized by ${{\mathbf{F}}_{\mathrm{RF}}}{{\mathbf{F}}_{\mathrm{BB}}}={{\mathbf{F}}_{\mathrm{opt}}}$ when a fully-digital array architecture is adopted, and $\mathbf{F}_{\mathrm{opt}}$ represents the optimal precoder for fully-digital beamforming \cite{palomar2003joint,viamimo}:
\begin{equation}
{{\mathbf{F}}_{\mathrm{opt}}}={{\left[ \mathbf{V} \right]}_{:,1:{{N}_{\mathrm{s}}}}}\in {{\mathbb{C}}^{{{N}_{\mathrm{t}}}\times {{N}_{\mathrm{s}}}}},
\label{eq6}
\end{equation}
where $\mathbf{V}$ is the right singular matrix of the channel's SVD, $\mathbf{H}=\mathbf{U\Sigma }{{\mathbf{V}}^{H}}$, with the eigenvalues in $\mathbf{\Sigma }$ arranged in descending order. 

For a hybrid array architecture, the hybrid beamforming problem can be equivalently reformulated as the following approximation problem, which aims to approximate the optimal fully-digital precoder \cite{el2014spatially,yu2016alternating}:
\begin{equation}
  \begin{aligned}
    & \underset{{{\mathbf{F}}_{\mathrm{RF}}},{{\mathbf{F}}_{\mathrm{BB}}}}{\mathop{\text{minimize}}}\,\quad \left\| {{\mathbf{F}}_{\mathrm{opt}}}-{{\mathbf{F}}_{\mathrm{RF}}}{{\mathbf{F}}_{\mathrm{BB}}} \right\|_{F}^{2} \\ 
   & \operatorname{subject\ to}\quad \left| {{\left[ {{\mathbf{F}}_{\mathrm{RF}}} \right]}_{i,j}} \right|={1}/{\sqrt{{{N}_{\mathrm{t}}}},\ \forall i,j}\; \\ 
   & \quad \quad \quad \quad \ \ \ \ \,\left\| {{\mathbf{F}}_{\mathrm{RF}}}{{\mathbf{F}}_{\mathrm{BB}}} \right\|_{F}^{2}={{N}_{\mathrm{s}}}\ . \\ 
  \end{aligned}
  \label{eq7}
  \end{equation}
However, solving (\ref{eq7}) remains challenging due to the coupling between analog and digital precoders ${{\mathbf{F}}_{\mathrm{RF}}}{{\mathbf{F}}_{\mathrm{BB}}}$, and no general closed-form solution for the global optimum has been derived to date. In the following, we assume perfect channel state information and develop an efficient algorithm that addresses this challenge by effectively decoupling the analog and digital precoders.

\section{Hybrid Beamforming with Alternating Residual Error Elimination}
An ideal approach to solving (\ref{eq7}) is to decouple ${{\mathbf{F}}_{\mathrm{RF}}}$ and ${{\mathbf{F}}_{\mathrm{BB}}}$ from the coupled product formulation ${{\mathbf{F}}_{\mathrm{RF}}}{{\mathbf{F}}_{\mathrm{BB}}}$, and then optimize them alternately. Specifically, the optimal solution satisfies ${{\mathbf{F}}_{\mathrm{RF}}}{{\mathbf{F}}_{\mathrm{BB}}}\approx{{\mathbf{F}}_{\mathrm{opt}}}$. First, by fixing ${{\mathbf{F}}_{\mathrm{RF}}}$, ${{\mathbf{F}}_{\mathrm{BB}}}$ can be decoupled as
\begin{equation} 
{{\mathbf{F}}_{\mathrm{BB}}}={{\left( \mathbf{F}_{\mathrm{RF}}^{H}{{\mathbf{F}}_{\mathrm{RF}}} \right)}^{-1}}\mathbf{F}_{\mathrm{RF}}^{H}{{\mathbf{F}}_{\mathrm{opt}}}.
\label{eq8}
\end{equation} 
Next, by fixing ${{\mathbf{F}}_{\mathrm{BB}}}$, ${{\mathbf{F}}_{\mathrm{RF}}}$ is decoupled as 
\begin{equation}
{{\mathbf{F}}_{\mathrm{RF}}}=\frac{1}{\sqrt{{{N}_{\mathrm{t}}}}}\exp \left( \mathrm{j}\cdot\arg \left( {{\mathbf{F}}_{\mathrm{opt}}}\mathbf{F}_{\mathrm{BB}}^{H}{{\left( {{\mathbf{F}}_{\mathrm{BB}}}\mathbf{F}_{\mathrm{BB}}^{H} \right)}^{-1}} \right) \right),
\label{eq9}
\end{equation}
where $\mathrm{j}=\sqrt{-1}$. By iteratively optimizing ${{\mathbf{F}}_{\mathrm{BB}}}$ and ${{\mathbf{F}}_{\mathrm{RF}}}$ alternately, the solution will eventually approach the optimal solution to (\ref{eq7}). However, since ${{\mathbf{F}}_{\mathrm{BB}}}\in {{\mathbb{C}}^{N_{\mathrm{RF}}^{\mathrm{t}}\times {{N}_{\mathrm{s}}}}}$ is a row rank-deficient matrix (due to $N_{\mathrm{RF}}^{\mathrm{t}}>{{N}_{\mathrm{s}}}$), its right inverse, denoted as $\mathbf{F}_{\mathrm{BB}}^{H}{{\left( {{\mathbf{F}}_{\mathrm{BB}}}\mathbf{F}_{\mathrm{BB}}^{H} \right)}^{-1}}$, does not exist. This makes it difficult to fully decouple ${{\mathbf{F}}_{\mathrm{RF}}}$ from the coupled product formulation without introducing decoupling errors, thereby significantly complicating the optimization process.

Different from existing algorithms \cite{el2014spatially,sohrabi2016hybrid,yu2016alternating,sohrabi2017hybrid,ioushua2019family,lee2014hybrid,chen2016compressive,lin2019hybrid,yuan2023alternating,zhu2023max} that directly tackle the original problem (\ref{eq7}) for precoder decoupling and hybrid beamforming, we propose a method that decomposes (\ref{eq7}) into two low-dimensional subproblems. Each subproblem independently decouples ${{\mathbf{F}}_{\mathrm{RF}}}$ and ${{\mathbf{F}}_{\mathrm{BB}}}$ from ${{\mathbf{F}}_{\mathrm{RF}}}{{\mathbf{F}}_{\mathrm{BB}}}$ with negligible decoupling errors. These subproblems alternately eliminate each other's residual error of the optimization result, enabling the original problem (\ref{eq7}) to converge toward the optimal solution. In the following, we introduce the proposed alternating residual error elimination (AREE) algorithm and analyze its performance.

\subsection{Alternating Residual Error Elimination Algorithm}
\begin{figure}[t]
  \centering
  \includegraphics[width=0.9\columnwidth]{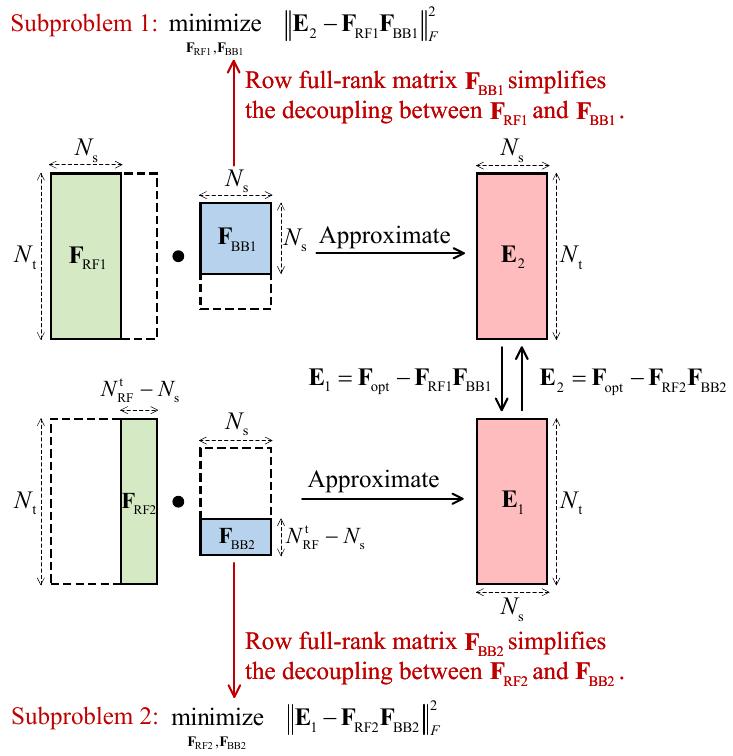}
  \caption{Diagram of the AREE algorithm.}
  \label{fig2}
\end{figure}
As illustrated in Fig. \ref{fig2}, we introduce an efficient and effective precoder decoupling method for hybrid beamforming based on the proposed AREE algorithm. The matrices ${{\mathbf{F}}_{\mathrm{RF}}}\in {{\mathbb{C}}^{{{N}_{\mathrm{t}}}\times N_{\mathrm{RF}}^{\mathrm{t}}}}$ and ${{\mathbf{F}}_{\mathrm{BB}}}\in {{\mathbb{C}}^{N_{\mathrm{RF}}^{\mathrm{t}}\times {{N}_{\mathrm{s}}}}}$ are each decomposed into two submatrices, namely ${\mathbf{F}}_{\mathrm{RF1}}$, ${\mathbf{F}}_{\mathrm{BB1}}$, ${\mathbf{F}}_{\mathrm{RF2}}$, and ${\mathbf{F}}_{\mathrm{BB2}}$ as follows:
\begin{equation}
   {{\mathbf{F}}_{\mathrm{RF1}}}={{\left[ {{\mathbf{F}}_{\mathrm{RF}}} \right]}_{:,1:{{N}_{\mathrm{s}}}}}\in {{\mathbb{C}}^{{{N}_{\mathrm{t}}}\times {{N}_{\mathrm{s}}}}},
\label{eq10}
\end{equation}
\begin{equation}
   {{\mathbf{F}}_{\mathrm{BB1}}}={{\left[ {{\mathbf{F}}_{\mathrm{BB}}} \right]}_{1:{{N}_{\mathrm{s}}},:}}\in {{\mathbb{C}}^{{{N}_{\mathrm{s}}}\times {{N}_{\mathrm{s}}}}}, 
\label{eq11}
\end{equation}
\begin{equation}
{{\mathbf{F}}_{\mathrm{RF2}}}={{\left[ {{\mathbf{F}}_{\mathrm{RF}}} \right]}_{:,{{N}_{\mathrm{s}}}+1:N_{\mathrm{RF}}^{\mathrm{t}}}}\in {{\mathbb{C}}^{{{N}_{\mathrm{t}}}\times \left( N_{\mathrm{RF}}^{\mathrm{t}}-{{N}_{\mathrm{s}}} \right)}},
\label{eq12}
\end{equation}
\begin{equation}
{{\mathbf{F}}_{\mathrm{BB2}}}={{\left[ {{\mathbf{F}}_{\mathrm{BB}}} \right]}_{{{N}_{\mathrm{s}}}+1:N_{\mathrm{RF}}^{\mathrm{t}},:}}\in {{\mathbb{C}}^{\left( N_{\mathrm{RF}}^{\mathrm{t}}-{{N}_{\mathrm{s}}} \right)\times {{N}_{\mathrm{s}}}}}.
\label{eq13}
\end{equation}
Accordingly, the original problem (\ref{eq7}) is decomposed into two subproblems. In the first subproblem, we optimize ${{\mathbf{F}}_{\mathrm{RF1}}}{{\mathbf{F}}_{\mathrm{BB1}}}$ to eliminate the residual error from the second subproblem. Then, in the second subproblem, we optimize ${{\mathbf{F}}_{\mathrm{RF2}}}{{\mathbf{F}}_{\mathrm{BB2}}}$ to eliminate the residual error from the first subproblem. By alternately optimizing these two subproblems, the original problem is eventually optimized. Since both ${{\mathbf{F}}_{\mathrm{BB1}}}$ and ${{\mathbf{F}}_{\mathrm{BB2}}}$ are row full-rank matrices, their right inverses exist, allowing the AREE algorithm to efficiently decouple ${{\mathbf{F}}_{\mathrm{RF1}}}$ and ${{\mathbf{F}}_{\mathrm{RF2}}}$ from ${{\mathbf{F}}_{\mathrm{RF1}}}{{\mathbf{F}}_{\mathrm{BB1}}}$ and ${{\mathbf{F}}_{\mathrm{RF2}}}{{\mathbf{F}}_{\mathrm{BB2}}}$ respectively with negligible decoupling errors. This not only reduces computational complexity but also suppresses decoupling errors, thereby enabling effective optimization of each RF chain.

\subsubsection{Optimization of the First Subproblem}
Firstly, we decompose (\ref{eq7}) into the first subproblem described in (\ref{eq14}), which aims to optimize the first ${{N}_{\mathrm{s}}}$ RF chains of ${{\mathbf{F}}_{\mathrm{RF}}}$:
\begin{equation}
\begin{aligned}
  & \underset{{{\mathbf{F}}_{\mathrm{RF1}}},{{\mathbf{F}}_{\mathrm{BB1}}}}{\mathop{\operatorname{minimize}}}\,\quad \left\| {{\mathbf{E}}_{\mathrm{2}}}-{{\mathbf{F}}_{\mathrm{RF1}}}{{\mathbf{F}}_{\mathrm{BB1}}} \right\|_{F}^{2} \\ 
 & \operatorname{subject\ to} \ \ \left| {{\left[ {{\mathbf{F}}_{\mathrm{RF1}}} \right]}_{i,j}} \right|={1}/{\sqrt{{{N}_{\mathrm{t}}}},\ \forall i,j}\;, \\ 
\end{aligned}
\label{eq14}
\end{equation}
where ${{\mathbf{E}}_{\mathrm{2}}}$ represents the residual error from (\ref{eq21}), and it is initialized as ${\mathbf{E}}_{\mathrm{2}}={{\mathbf{F}}_{\mathrm{opt}}}$ for the first optimization of (\ref{eq14}). The power constraint $\left\| {{\mathbf{F}}_{\mathrm{RF1}}}{{\mathbf{F}}_{\mathrm{BB1}}} \right\|_{F}^{2}<{{N}_{\mathrm{s}}}$ is temporarily omitted here, and we will later demonstrate that this does not affect the correctness of the result. 

First, by fixing ${{\mathbf{F}}_{\mathrm{RF1}}}$, ${{\mathbf{F}}_{\mathrm{BB1}}}$ is treated as the sole optimization variable. The derivative of the objective function is expressed as: 
\begin{equation}
  \frac{\partial \left\| {{\mathbf{E}}_{\mathrm{2}}}-{{\mathbf{F}}_{\mathrm{RF1}}}{{\mathbf{F}}_{\mathrm{BB1}}} \right\|_{F}^{2}}{\partial {{\mathbf{F}}_{\mathrm{BB1}}}}=\mathbf{0}\ ,
  \label{eq15}
\end{equation}
where the optimal solution for ${{\mathbf{F}}_{\mathrm{BB1}}}$ is then calculated as
\begin{equation}
  {{\mathbf{F}}_{\mathrm{BB1}}}=\mathbf{F}_{\mathrm{RF1}}^{\dagger }{{\mathbf{E}}_{\mathrm{2}}}\ .
  \label{eq16}
\end{equation}
Next, by fixing ${{\mathbf{F}}_{\mathrm{BB1}}}$, ${{\mathbf{F}}_{\mathrm{RF1}}}$ is treated as the sole optimization variable. Since both ${{\mathbf{E}}_{\mathrm{2}}}$ and ${{\mathbf{F}}_{\mathrm{BB1}}}$ are not semi-unitary matrices, deriving the optimal solution for ${{\mathbf{F}}_{\mathrm{RF1}}}$ of (\ref{eq14}) becomes challenging. However, since ${{\mathbf{F}}_{\mathrm{BB1}}}$ is a row full-rank matrix (for which the right inverse exists), a suboptimal solution for ${{\mathbf{F}}_{\mathrm{RF1}}}$ can be derived: by temporarily ignoring the constant modulus constraint $\left| {{\left[ {{\mathbf{F}}_{\mathrm{RF1}}} \right]}_{i,j}} \right|={1}/{\sqrt{{{N}_{\mathrm{t}}}}}\;$ in (\ref{eq14}), we calculate the derivative of the objective function with respect to ${{\mathbf{F}}_{\mathrm{RF1}}}$ as follows:
\begin{equation}
\begin{aligned}
  & \quad \ \frac{\partial \left\| {{\mathbf{E}}_{\mathrm{2}}}-{{\mathbf{F}}_{\mathrm{RF1}}}{{\mathbf{F}}_{\mathrm{BB1}}} \right\|_{F}^{2}}{\partial {{\mathbf{F}}_{\mathrm{RF1}}}} \\ 
 & =\partial \operatorname{Tr}\left( \mathbf{E}_{\mathrm{2}}^{H}{{\mathbf{E}}_{\mathrm{2}}}-\mathbf{E}_{\mathrm{2}}^{H}{{\mathbf{F}}_{\mathrm{RF1}}}{{\mathbf{F}}_{\mathrm{BB1}}}-\mathbf{F}_{\mathrm{BB1}}^{H}\mathbf{F}_{\mathrm{RF1}}^{H}{{\mathbf{E}}_{\mathrm{2}}} \right. \\ 
 & {\left. \quad \quad \quad \quad \quad \quad \ \ \ \ +\mathbf{F}_{\mathrm{BB1}}^{H}\mathbf{F}_{\mathrm{RF1}}^{H}{{\mathbf{F}}_{\mathrm{RF1}}}{{\mathbf{F}}_{\mathrm{BB1}}} \right)}/{\partial {{\mathbf{F}}_{\mathrm{RF1}}}}\; \\ 
 & =-\mathbf{E}_{\mathrm{2}}^{*}\mathbf{F}_{\mathrm{BB1}}^{T}+\mathbf{F}_{\mathrm{RF1}}^{*}\mathbf{F}_{\mathrm{BB1}}^{*}\mathbf{F}_{\mathrm{BB1}}^{T} \ .\\
\end{aligned}
\label{eq17}
\end{equation}
Setting the derivative to zero, the unconstrained optimal solution is calculated as ${{\mathbf{F}}_{\mathrm{RF1}}}=\mathbf{E}_{\mathrm{2}}{\mathbf{F}}_{\mathrm{BB1}}^{H}{{\left( {{\mathbf{F}}_{\mathrm{BB1}}}\mathbf{F}_{\mathrm{BB1}}^{H} \right)}^{\dagger }}$. This reformulates (\ref{eq14}) into a constant modulus constrained optimization problem:
\begin{equation}
\begin{aligned}
  & \underset{{{\mathbf{F}}_{\mathrm{RF1}}}}{\mathop{\operatorname{minimize}}}\,\quad \left\| {\mathbf{E}_\mathrm{2}}{\mathbf{F}}_{\mathrm{BB1}}^{H}{{\left( {{\mathbf{F}}_{\mathrm{BB1}}}\mathbf{F}_{\mathrm{BB1}}^{H} \right)}^{\dagger }}-{{\mathbf{F}}_{\mathrm{RF1}}} \right\|_{F}^{2} \\ 
 & \operatorname{subject\ to}\quad \left| {{\left[ {{\mathbf{F}}_{\mathrm{RF1}}} \right]}_{i,j}} \right|={1}/{\sqrt{{{N}_{\mathrm{t}}}},\ \forall i,j}\; . \\ 
\end{aligned}
\label{eq18}
\end{equation}
The optimal solution for (\ref{eq18}) is then given by 
\begin{equation}
  {{\mathbf{F}}_{\mathrm{RF1}}}=\frac{1}{\sqrt{{{N}_{\mathrm{t}}}}}\exp \left( \mathrm{j}\cdot \arg \left( {\mathbf{E}_{\mathrm{2}}}{\mathbf{F}}_{\mathrm{BB1}}^{H}{{\left( {{\mathbf{F}}_{\mathrm{BB1}}}\mathbf{F}_{\mathrm{BB1}}^{H} \right)}^{\dagger }} \right) \right).
  \label{eq19}
\end{equation}
Therefore, (\ref{eq19}) serves as a suboptimal solution for subproblem (\ref{eq14}). Consequently, ${{\mathbf{F}}_{\mathrm{BB1}}}$ and ${{\mathbf{F}}_{\mathrm{RF1}}}$ are both effectively decoupled from product formulation in (\ref{eq14}). By alternating iterations of (\ref{eq16}) and (\ref{eq19}), the first subproblem (\ref{eq14}) converges to a suboptimal solution, that is, $\mathbf{F}_{\mathrm{BB1}}^{\mathrm{opt}}$ and $\mathbf{F}_{\mathrm{RF1}}^{\mathrm{opt}}$. The residual error after optimizing (\ref{eq14}) is eventually expressed as
\begin{equation}
  {\mathbf{E}_{\mathrm{1}}}={{\mathbf{F}}_{\mathrm{opt}}}-\mathbf{F}_{\mathrm{RF1}}^{\mathrm{opt}}\mathbf{F}_{\mathrm{BB1}}^{\mathrm{opt}}\ .
  \label{eq20}
\end{equation}

\subsubsection{Optimization of the Second Subproblem}
Next, we decompose (\ref{eq7}) into the second subproblem described in (\ref{eq21}), which is designed to optimize the remaining $N_{\mathrm{RF}}^{\mathrm{t}}-{{N}_{\mathrm{s}}}$ RF chains in ${{\mathbf{F}}_{\mathrm{RF}}}$ by eliminating the residual error ${\mathbf{E}_{\mathrm{1}}}$ from the first subproblem (\ref{eq14}): 
\begin{equation}
\begin{aligned}
  & \underset{{{\mathbf{F}}_{\mathrm{RF2}}},{{\mathbf{F}}_{\mathrm{BB2}}}}{\mathop{\operatorname{minimize}}}\,\quad \left\| {\mathbf{E}_{\mathrm{1}}}-{{\mathbf{F}}_{\mathrm{RF2}}}{{\mathbf{F}}_{\mathrm{BB2}}} \right\|_{F}^{2} \\ 
 & {\operatorname{subject\ to}}\quad \left| {{\left[ {{\mathbf{F}}_{\mathrm{RF2}}} \right]}_{i,j}} \right|={1}/{\sqrt{{{N}_{\mathrm{t}}}},\ \forall i,j}\; , \\ 
\end{aligned}
\label{eq21}
\end{equation}
where the power constraint $\left\| {{\mathbf{F}}_{\mathrm{RF2}}}{{\mathbf{F}}_{\mathrm{BB2}}} \right\|_{F}^{2}<{{N}_{\mathrm{s}}}$ is temporarily omitted here, and we will later demonstrate that this does not affect the result's correctness. The optimal solution for ${{\mathbf{F}}_{\mathrm{BB2}}}$ and suboptimal solution for ${{\mathbf{F}}_{\mathrm{RF2}}}$ in (\ref{eq21}) can be calculated in a manner similar to processes in (\ref{eq16}) and (\ref{eq19}), respectively:
\begin{equation}
  {{\mathbf{F}}_{\mathrm{BB2}}}=\mathbf{F}_{\mathrm{RF2}}^{\dagger }{\mathbf{E}_{\mathrm{1}}}\ ,  
  \label{eq22}
\end{equation}
\begin{equation}
  {{\mathbf{F}}_{\mathrm{RF2}}}=\frac{1}{\sqrt{{{N}_{\mathrm{t}}}}}\exp \left( \mathrm{j}\cdot \arg \left( {\mathbf{E}_{\mathrm{1}}}{\mathbf{F}}_{\mathrm{BB2}}^{H}{{\left( {{\mathbf{F}}_{\mathrm{BB2}}}\mathbf{F}_{\mathrm{BB2}}^{H} \right)}^{\dagger }} \right) \right).
  \label{eq23}
\end{equation}
Therefore, ${{\mathbf{F}}_{\mathrm{BB2}}}$ and ${{\mathbf{F}}_{\mathrm{RF2}}}$ are both effectively decoupled from product formulation in (\ref{eq21}). By alternating iterations of (\ref{eq22}) and (\ref{eq23}), the second subproblem (\ref{eq21}) converges to a suboptimal solution, that is, $\mathbf{F}_{\mathrm{BB2}}^{\mathrm{opt}}$ and $\mathbf{F}_{\mathrm{RF2}}^{\mathrm{opt}}$. Subsequently, the residual error of the optimization result in (\ref{eq21}) relative to ${{\mathbf{F}}_{\mathrm{opt}}}$ is updated as
\begin{equation}
{\mathbf{E}_{\mathrm{2}}}={{\mathbf{F}}_{\mathrm{opt}}}-\mathbf{F}_{\mathrm{RF2}}^{\mathrm{opt}}\mathbf{F}_{\mathrm{BB2}}^{\mathrm{opt}},
\label{eq24}
\end{equation} 
where $\mathbf{E}_{\mathrm{2}}$ is then fed back into the first subproblem (\ref{eq14}) to re-optimize and update $\mathbf{F}_{\mathrm{BB1}}^{\mathrm{opt}}$ and $\mathbf{F}_{\mathrm{RF1}}^{\mathrm{opt}}$.

The AREE algorithm achieves a suboptimal solution through an iterative feedback process: the residual error $\mathbf{E}_{\mathrm{2}}$ from (\ref{eq21}) is fed back into (\ref{eq14}), and then the residual error ${{\mathbf{E}}_{\mathrm{1}}}$ from (\ref{eq14}) is fed back into (\ref{eq21}). By alternating feedback between (\ref{eq14}) and (\ref{eq21}) iteratively until the changes of $\mathbf{E}_{\mathrm{1}}$ and ${{\mathbf{E}}_{\mathrm{2}}}$ fall below a specified threshold, the results $\mathbf{F}_{\mathrm{RF1}}^{\mathrm{opt}}$, $\mathbf{F}_{\mathrm{BB1}}^{\mathrm{opt}}$, $\mathbf{F}_{\mathrm{RF2}}^{\mathrm{opt}}$ and $\mathbf{F}_{\mathrm{BB2}}^{\mathrm{opt}}$ are effectively optimized.

\subsubsection{Combination of the Results under Power Constraint}
Without considering the power constraint $\left\| \mathbf{F}_{\mathrm{RF}}\mathbf{F}_{\mathrm{BB}} \right\|_{F}^{2}={{N}_{\mathrm{s}}}$, the suboptimal solutions for the original hybrid beamforming problem (\ref{eq7}) can be constructed as follows:
\begin{equation}
\mathbf{F}_{\mathrm{RF}}^{\mathrm{opt}}=\left[ \mathbf{F}_{\mathrm{RF1}}^{\mathrm{opt}},\ \mathbf{F}_{\mathrm{RF2}}^{\mathrm{opt}} \right]\in {{\mathbb{C}}^{{{N}_{\mathrm{t}}}\times N_{\mathrm{RF}}^{\mathrm{t}}}},
\label{eq25}
\end{equation}
\begin{equation}
\mathbf{F}_{\mathrm{BB}}^{\mathrm{opt}}=\left[ \begin{aligned}
  & \mathbf{F}_{\mathrm{BB1}}^{\mathrm{opt}} \\ 
 & \mathbf{F}_{\mathrm{BB2}}^{\mathrm{opt}} \\ 
\end{aligned} \right]\in {{\mathbb{C}}^{N_{\mathrm{RF}}^{\mathrm{t}}\times {{N}_{\mathrm{s}}}}}.
\label{eq26}
\end{equation}
Taking the power constraint in (\ref{eq7}) into consideration, (\ref{eq26}) should be modified into the following power-normalized form:
\begin{equation}
  \mathbf{\hat{F}}_{\mathrm{BB}}^{\mathrm{opt}}=\frac{\sqrt{{{N}_{\mathrm{s}}}}}{{{\left\| \mathbf{F}_{\mathrm{RF}}^{\mathrm{opt}}\mathbf{F}_{\mathrm{BB}}^{\mathrm{opt}} \right\|}_{F}}}\mathbf{F}_{\mathrm{BB}}^{\mathrm{opt}}\ .
  \label{eq27}
\end{equation}
This indicates that the final solution {\small $\mathbf{\hat{F}}_{\mathrm{BB}}^{\mathrm{opt}}$} is not directly composed of {\small $\mathbf{F}_{\mathrm{BB1}}^{\mathrm{opt}}$} and {\small $\mathbf{F}_{\mathrm{BB2}}^{\mathrm{opt}}$}, but their power-normalized counterparts: {\small $\mathbf{\hat{F}}_{\mathrm{BB1}}^{\mathrm{opt}}=\left( {\sqrt{{{N}_{\mathrm{s}}}}}/{{{\left\| \mathbf{F}_{\mathrm{RF}}^{\mathrm{opt}}\mathbf{F}_{\mathrm{BB}}^{\mathrm{opt}} \right\|}_{F}}}\; \right)\mathbf{F}_{\mathrm{BB1}}^{\mathrm{opt}}$} and {\small $\mathbf{\hat{F}}_{\mathrm{BB2}}^{\mathrm{opt}}=\left( {\sqrt{{{N}_{\mathrm{s}}}}}/{{{\left\| \mathbf{F}_{\mathrm{RF}}^{\mathrm{opt}}\mathbf{F}_{\mathrm{BB}}^{\mathrm{opt}} \right\|}_{F}}}\; \right)\mathbf{F}_{\mathrm{BB2}}^{\mathrm{opt}}$}. We demonstrate in the \textbf{Appendix} that if {\small $\mathbf{F}_{\mathrm{BB1}}^{\mathrm{opt}}$} and {\small $\mathbf{F}_{\mathrm{BB2}}^{\mathrm{opt}}$} approximate the suboptimal solutions of (\ref{eq14}) and (\ref{eq21}), then {\small $\mathbf{\hat{F}}_{\mathrm{BB1}}^{\mathrm{opt}}$} and {\small $\mathbf{\hat{F}}_{\mathrm{BB2}}^{\mathrm{opt}}$} will also approach them with similar approximation errors. In other words, first neglecting the power constraints in (\ref{eq14}) and (\ref{eq21}) and then applying them to the results after the optimization is complete, does not compromise the correctness of results. The AREE algorithm is summarized in \textbf{Algorithm 1}, where the initial values $\mathbf{F}_{\mathrm{RF}}^{\mathrm{initial}}$ and $\mathbf{F}_{\mathrm{BB}}^{\mathrm{initial}}$ can be initialized randomly.\hfill$\blacksquare$

\subsection{Convergence Analysis}
The objective function of (\ref{eq7}) after the $t\text{-th}$ iteration can be expressed as
\begin{equation}
  {{f}^{\left( t \right)}}=\left\| {{\mathbf{F}}_{\text{opt}}}-\mathbf{F}_{\text{RF1}}^{\left( t \right)}\mathbf{F}_{\text{BB1}}^{\left( t \right)}-\mathbf{F}_{\text{RF2}}^{\left( t \right)}\mathbf{F}_{\text{BB2}}^{\left( t \right)} \right\|_{F}^{2}.
\end{equation}
During the $\left( t+1 \right)\text{-th}$ iteration, $\mathbf{F}_{\text{RF1}}^{\left( t \right)}\mathbf{F}_{\text{BB1}}^{\left( t \right)}$ is updated to $\mathbf{F}_{\text{RF1}}^{\left( t+1 \right)}\mathbf{F}_{\text{BB1}}^{\left( t+1 \right)}$ after optimizing the first subproblem (\ref{eq14}). Consequently, ${{f}^{\left( t \right)}}$ is updated to the intermediate objective ${{f}^{\left( t+{1}/{2}\; \right)}}$ before optimizing the second subproblem (\ref{eq21}):
\begin{equation}
  \begin{aligned}
  & {{f}^{\left( t+{1}/{2}\; \right)}}=\left\| {{\mathbf{F}}_{\text{opt}}}-\mathbf{F}_{\text{RF1}}^{\left( t+1 \right)}\mathbf{F}_{\text{BB1}}^{\left( t+1 \right)}-\mathbf{F}_{\text{RF2}}^{\left( t \right)}\mathbf{F}_{\text{BB2}}^{\left( t \right)} \right\|_{F}^{2} \\ 
 & =\left\| \mathbf{E}_{2}^{\left( t \right)}-\mathbf{F}_{\text{RF1}}^{\left( t+1 \right)}\mathbf{F}_{\text{BB1}}^{\left( t+1 \right)} \right\|_{F}^{2}<\left\| \mathbf{E}_{2}^{\left( t \right)}-\mathbf{F}_{\text{RF1}}^{\left( t \right)}\mathbf{F}_{\text{BB1}}^{\left( t \right)} \right\|_{F}^{2}={{f}^{\left( t \right)}}, \\ 
\end{aligned}
\label{eq_new29}
\end{equation}
where the inequality holds because (\ref{eq14}) has been optimized. Subsequently, $\mathbf{F}_{\text{RF2}}^{\left( t \right)}\mathbf{F}_{\text{BB2}}^{\left( t \right)}$ is updated to $\mathbf{F}_{\text{RF2}}^{\left( t+1 \right)}\mathbf{F}_{\text{BB2}}^{\left( t+1 \right)}$ after optimizing the second subproblem (\ref{eq21}), completing the $\left( t+1 \right)\text{-th}$ iteration. The objective ${{f}^{\left( t+{1}/{2}\; \right)}}$ is then updated to ${{f}^{\left( t+1 \right)}}$:
\begin{equation}
  \begin{aligned}
  & {{f}^{\left( t+1 \right)}}=\left\| {{\mathbf{F}}_{\text{opt}}}-\mathbf{F}_{\text{RF1}}^{\left( t+1 \right)}\mathbf{F}_{\text{BB1}}^{\left( t+1 \right)}-\mathbf{F}_{\text{RF2}}^{\left( t+1 \right)}\mathbf{F}_{\text{BB2}}^{\left( t+1 \right)} \right\|_{F}^{2} \\ 
 & =\!\left\| \mathbf{E}_{1}^{\left( t+\!1 \right)}\!\!-\!\mathbf{F}_{\text{RF2}}^{\left( t+\!1 \right)}\mathbf{F}_{\text{BB2}}^{\left( t+\!1 \right)} \right\|_{F}^{2}\!\!<\!\left\| \mathbf{E}_{1}^{\left( t+\!1 \right)}\!\!-\!\mathbf{F}_{\text{RF2}}^{\left( t \right)}\mathbf{F}_{\text{BB2}}^{\left( t \right)} \right\|_{F}^{2}\!\!=\!\!{{f}^{\left( t+{1}\!/{2} \right)}}, \\ 
\end{aligned}
\label{eq_new30}
\end{equation}
where the inequality holds because (\ref{eq21}) has been optimized. Therefore, it follows from (\ref{eq_new29}) and (\ref{eq_new30}) that ${{f}^{\left( t+1 \right)}}<{{f}^{\left( t+{1}/{2}\!\; \right)}}<{{f}^{\left( t \right)}}$, and since it is lower-bounded by a non-negative value, i.e., ${{f}^{\left( t \right)}}>0$, the AREE algorithm is guaranteed to converge, as also validated by the simulation results in Sec. VII.

\subsection{Condition for the Optimal Solution of the First Subproblem}
\subsubsection{Optimality Condition for the First Subproblem}
For the first subproblem (\ref{eq14}), since $\mathbf{F}_{\mathrm{BB1}}$ is a square matrix satisfying $\mathbf{F}_{\mathrm{BB1}}^{H}{{\left( {{\mathbf{F}}_{\mathrm{BB1}}}\mathbf{F}_{\mathrm{BB1}}^{H} \right)}^{\dagger }}=\mathbf{F}_{\mathrm{BB1}}^{-1}$, the bounds of the objective function can be derived using Cauchy-Schwarz inequality:
\begin{equation}
\begin{aligned}
  & \,\quad \left\| {{\mathbf{E}}_{2}}\mathbf{F}_{\mathrm{BB1}}^{H}{{\left( {{\mathbf{F}}_{\mathrm{BB1}}}\mathbf{F}_{\mathrm{BB1}}^{H} \right)}^{\dagger }}-{{\mathbf{F}}_{\mathrm{RF1}}} \right\|_{F}^{2}{{\left( \left\| \mathbf{F}_{\mathrm{BB1}}^{-1} \right\|_{F}^{2} \right)}^{-1}} \\ 
 & \le \left\| {{\mathbf{E}}_{2}}-{{\mathbf{F}}_{\mathrm{RF1}}}{{\mathbf{F}}_{\mathrm{BB1}}} \right\|_{F}^{2} \\ 
 & \le \left\| {{\mathbf{E}}_{2}}\mathbf{F}_{\mathrm{BB1}}^{H}{{\left( {{\mathbf{F}}_{\mathrm{BB1}}}\mathbf{F}_{\mathrm{BB1}}^{H} \right)}^{\dagger }}-{{\mathbf{F}}_{\mathrm{RF1}}} \right\|_{F}^{2}\left\| {{\mathbf{F}}_{\mathrm{BB1}}} \right\|_{F}^{2}, \\ 
\end{aligned}
\label{eq28}
\end{equation}
which indicates that subproblem (\ref{eq14}) will converge as (\ref{eq18}) is optimized. Specifically, when $\mathbf{F}_{\mathrm{BB1}}$ is a unitary matrix, (\ref{eq19}) becomes the optimal solution to (\ref{eq14}), because the objective function of (\ref{eq14}) can then be rewritten as 
\begin{equation}
\begin{aligned}
  & \quad  \left\| {{\mathbf{E}}_{2}}-{{\mathbf{F}}_{\mathrm{RF1}}}{{\mathbf{F}}_{\mathrm{BB1}}} \right\|_{F}^{2} \\ 
 & =\operatorname{Tr}\left( \mathbf{E}_{2}^{H}{{\mathbf{E}}_{2}} \right)-\operatorname{Tr}\left( \mathbf{E}_{2}^{H}{{\mathbf{F}}_{\mathrm{RF1}}}{{\mathbf{F}}_{\mathrm{BB1}}} \right)-\operatorname{Tr}\left( \mathbf{F}_{\mathrm{BB1}}^{H}\mathbf{F}_{\mathrm{RF1}}^{H}{{\mathbf{E}}_{2}} \right) \\ 
 & \quad \ +\operatorname{Tr}\left( \mathbf{F}_{\mathrm{BB1}}^{H}\mathbf{F}_{\mathrm{RF1}}^{H}{{\mathbf{F}}_{\mathrm{RF1}}}{{\mathbf{F}}_{\mathrm{BB1}}} \right) \\ 
 & =\operatorname{Tr}\left( {{\mathbf{F}}_{\mathrm{BB1}}}\mathbf{E}_{2}^{H}{{\mathbf{E}}_{2}}\mathbf{F}_{\mathrm{BB1}}^{H} \right)-\operatorname{Tr}\left( \mathbf{E}_{2}^{H}{{\mathbf{F}}_{\mathrm{RF1}}}{{\mathbf{F}}_{\mathrm{BB1}}} \right) \\
 & \quad \ -\operatorname{Tr}\left( \mathbf{F}_{\mathrm{BB1}}^{H}\mathbf{F}_{\mathrm{RF1}}^{H}{{\mathbf{E}}_{2}} \right)+\operatorname{Tr}\left( \mathbf{F}_{\mathrm{RF1}}^{H}{{\mathbf{F}}_{\mathrm{RF1}}} \right) \\ 
 & =\left\| {{\mathbf{E}}_{2}}\mathbf{F}_{\mathrm{BB1}}^{H}-{{\mathbf{F}}_{\mathrm{RF1}}} \right\|_{F}^{2}, \\ 
\end{aligned}
\label{eq29}
\end{equation}
which is identical to (\ref{eq18}). Therefore, (\ref{eq19}) becomes the optimal solution to (\ref{eq14}). 

\subsubsection{Evaluation of the Optimality Condition}
As shown in (\ref{eq29}), (\ref{eq19}) becomes the optimal solution to (\ref{eq14}) when $\mathbf{F}_{\mathrm{BB1}}$ is a unitary matrix. To evaluate the deviation of $\mathbf{F}_{\mathrm{BB1}}$ from a unitary matrix during the AREE algorithm, we define the normalized mean squared error (NMSE) as: 
\begin{equation}
\text{NMSE}\!\left( {{\mathbf{F}}_{\mathrm{\!BB1}}} \!\right)\!=\!{\left\| \frac{{{\mathbf{F}}_{\mathrm{\!BB1}}}\mathbf{F}_{\mathrm{\!BB1}}^{H}}{{{\left\| {{\mathbf{F}}_{\mathrm{\!BB1}}}\mathbf{F}_{\mathrm{\!BB1}}^{H} \right\|}_{F}}}\!-\!\frac{{{\mathbf{I}}_{{{N}_{\mathrm{s}}}}}}{{{\left\| {{\mathbf{I}}_{{{N}_{\mathrm{s}}}}} \right\|}_{F}}} \right\|_{F}^{2}}/{\left\| \frac{{{\mathbf{I}}_{{{N}_{\mathrm{s}}}}}}{{{\left\| {{\mathbf{I}}_{{{N}_{\mathrm{s}}}}} \right\|}_{F}}} \right\|_{F}^{2}}\;.
\label{eq30}
\end{equation}
As demonstrated in~Sec. VII, $\mathbf{F}_{\mathrm{BB1}}$ approaches a unitary matrix during the optimization process of the AREE algorithm, making (\ref{eq19}) nearly optimal. 

Similarly, for the second subproblem (\ref{eq21}), as ${N}_{\mathrm{RF}}^{\mathrm{t}}$ increases, $\mathbf{F}_{\mathrm{BB2}}$ approximates a square matrix. Consequently, the second subproblem (\ref{eq21}) becomes more similar to the first subproblem (\ref{eq14}). As shown in~Sec. VII, $\mathbf{F}_{\mathrm{BB2}}$ approaches a semi-unitary matrix during AREE algorithm, thus enabling (\ref{eq23}) to converge to the optimal solution as ${N}_{\mathrm{RF}}^{\mathrm{t}}$ increases. When the number of RF chains approaches $\mathrm{2}{N}_{\mathrm{s}}$, the AREE algorithm achieves the fully-digital optimal performance.

\begin{algorithm}[t]
  \caption{AREE hybrid beamforming algorithm}\label{algorithm2}
  \begin{algorithmic}[1]
      \Require{${{\mathbf{F}}_{\mathrm{opt}}}$, $\mathbf{F}_{\mathrm{RF}}^{\mathrm{initial}}$, $\mathbf{F}_{\mathrm{BB}}^{\mathrm{initial}}$, $N_{\mathrm{RF}}^{\mathrm{t}}$, ${{N}_{\mathrm{s}}}$}
      \State
      \textbf{Initialization:} 
      \State
      \quad \ ${{\mathbf{E}}_{\mathrm{2}}}={{\mathbf{F}}_{\mathrm{opt}}}$, ${{\mathbf{F}}_{\mathrm{RF}}}=\mathbf{F}_{\mathrm{RF}}^{\mathrm{initial}}$, ${{\mathbf{F}}_{\mathrm{BB}}}=\mathbf{F}_{\mathrm{BB}}^{\mathrm{initial}}$, 
      \State 
      \quad \ ${{\mathbf{F}}_{\mathrm{RF1}}}={{\left[ {{\mathbf{F}}_{\mathrm{RF}}} \right]}_{:,1:{{N}_{\mathrm{s}}}}}$, ${{\mathbf{F}}_{\mathrm{RF2}}}={{\left[ {{\mathbf{F}}_{\mathrm{RF}}} \right]}_{:,{{N}_{\mathrm{s}}}+1:N_{\mathrm{RF}}^{\mathrm{t}}}}$,
      \State
      \quad \ ${{\mathbf{F}}_{\mathrm{BB1}}}={{\left[ {{\mathbf{F}}_{\mathrm{BB}}} \right]}_{1:{{N}_{\mathrm{s}}},:}}$, ${{\mathbf{F}}_{\mathrm{BB2}}}={{\left[ {{\mathbf{F}}_{\mathrm{BB}}} \right]}_{{{N}_{\mathrm{s}}}+1:N_{\mathrm{RF}}^{\mathrm{t}},:}}$\ ;  
      \State
      \textbf{repeat:} (defined as $iterations3$)
      \State
      \quad \ \textbf{repeat:} (defined as $iterations1$)
      \State
      \quad \ \quad \ ${{\mathbf{F}}_{\mathrm{BB1}}}=\mathbf{F}_{\mathrm{RF1}}^{\dagger }{{\mathbf{E}}_{\mathrm{2}}}$,
      \State
        \quad \ \quad \ ${\mathbf{F}}_{\mathrm{\!RF1}}\!\!=\!\!\frac{1}{\sqrt{{{N}_{\mathrm{t}}}}}\!\exp \!\left( \mathrm{j}\!\cdot\! \arg \left( {{\mathbf{E}}_{\mathrm{2}}}\mathbf{F}_{\mathrm{\!\!B\!B1}}^{H}{{\left( {{\mathbf{F}}_{\mathrm{\!\!B\!B1}}}\mathbf{F}_{\mathrm{\!\!B\!B1}}^{H} \right)}^{\dagger }} \right) \right)$;
      \State
      \quad \ \textbf{until} a stopping criterion triggers
      \State
      \quad \ $\mathbf{F}_{\mathrm{RF1}}^{\mathrm{opt}}={{\mathbf{F}}_{\mathrm{RF1}}}$, $\mathbf{F}_{\mathrm{BB1}}^{\mathrm{opt}}={{\mathbf{F}}_{\mathrm{BB1}}}$,
      \State
      \quad \ ${\mathbf{E}_{\mathrm{1}}}={{\mathbf{F}}_{\mathrm{opt}}}-\mathbf{F}_{\mathrm{RF1}}^{\mathrm{opt}}\mathbf{F}_{\mathrm{BB1}}^{\mathrm{opt}}$;
      \State
      \quad \ \textbf{repeat:} (defined as $iterations2$)
      \State
      \quad \ \quad \ ${{\mathbf{F}}_{\mathrm{BB2}}}=\mathbf{F}_{\mathrm{RF2}}^{\dagger }{\mathbf{E}_{\mathrm{1}}}$,
      \State
      \quad \ \quad \ ${{\mathbf{F}}_{\mathrm{\!RF2}}}\!=\!\frac{1}{\sqrt{{{N}_{\mathrm{t}}}}}\!\exp \!\left( \mathrm{j}\!\cdot\! \arg \left( {\mathbf{E}_{\mathrm{1}}}{\mathbf{F}}_{\mathrm{\!\!B\!B2}}^{H}{{\left( {{\mathbf{F}}_{\mathrm{\!\!B\!B2}}}\mathbf{F}_{\mathrm{\!\!B\!B2}}^{H} \right)}^{\dagger }} \right) \right)$;
      \State
      \quad \ \textbf{until} a stopping criterion triggers
      \State
      \quad \ $\mathbf{F}_{\mathrm{RF2}}^{\mathrm{opt}}={{\mathbf{F}}_{\mathrm{RF2}}}$, $\mathbf{F}_{\mathrm{BB2}}^{\mathrm{opt}}={{\mathbf{F}}_{\mathrm{BB2}}}$,
      \State
      \quad \ ${{\mathbf{E}}_{\mathrm{2}}}={{\mathbf{F}}_{\mathrm{opt}}}-\mathbf{F}_{\mathrm{RF2}}^{\mathrm{opt}}\mathbf{F}_{\mathrm{BB2}}^{\mathrm{opt}}$;
      \State
      \textbf{until} a stopping criterion triggers
      \State
      $\mathbf{F}_{\mathrm{\!RF}}^{\mathrm{opt}}\!=\!\left[ \mathbf{F}_{\mathrm{RF1}}^{\mathrm{opt}},\ \mathbf{F}_{\mathrm{RF2}}^{\mathrm{opt}} \right]$, $\mathbf{F}_{\mathrm{\!BB}}^{\mathrm{opt}}\!=\!{{\left[ {{\left( \mathbf{F}_{\mathrm{\!B\!B1}}^{\mathrm{opt}} \right)}^{T}},\ \! {{\left( \mathbf{F}_{\mathrm{\!B\!B2}}^{\mathrm{opt}} \right)}^{T}} \right]}^{T}}$,
      \State
      $\mathbf{\hat{F}}_{\mathrm{BB}}^{\mathrm{opt}}=\frac{\sqrt{{{N}_{\mathrm{s}}}}}{{{\left\| \mathbf{F}_{\mathrm{RF}}^{\mathrm{opt}}\mathbf{F}_{\mathrm{BB}}^{\mathrm{opt}} \right\|}_{F}}}\mathbf{F}_{\mathrm{BB}}^{\mathrm{opt}}$;
      \Ensure
      $\mathbf{F}_{\mathrm{RF}}^{\mathrm{opt}}$, $\mathbf{\hat{F}}_{\mathrm{BB}}^{\mathrm{opt}}$
  \end{algorithmic}
\end{algorithm}

\subsection{The Optimal Matrix Partition of the Original Problem}
In the AREE algorithm, ${{\mathbf{F}}_{\mathrm{BB}}}$ is partitioned into a square matrix ${{\mathbf{F}}_{\mathrm{BB1}}}$ and a row full-rank matrix ${{\mathbf{F}}_{\mathrm{BB2}}}$, with ${{\mathbf{F}}_{\mathrm{RF}}}$ correspondingly partitioned to maintain the matrix multiplication relationship. In fact, ${{\mathbf{F}}_{\mathrm{BB}}}$ and ${{\mathbf{F}}_{\mathrm{RF}}}$ can have other partitioning schemes. For example, ${{\mathbf{F}}_{\mathrm{BB}}}$ can also be partitioned as
\begin{equation}
  {{\mathbf{\tilde{F}}}_{\mathrm{BB1}}}={{\left[ {{\mathbf{F}}_{\mathrm{BB}}} \right]}_{1:n,:}}\in {{\mathbb{C}}^{n\times {{N}_{\mathrm{s}}}}},
  \label{eq31}
\end{equation}
\begin{equation}
  {{\mathbf{\tilde{F}}}_{\mathrm{BB2}}}={{\left[ {{\mathbf{F}}_{\mathrm{BB}}} \right]}_{n+1:N_{\mathrm{RF}}^{\mathrm{t}},:}}\in {{\mathbb{C}}^{\left( N_{\mathrm{RF}}^{\mathrm{t}}-n \right)\times {{N}_{\mathrm{s}}}}},
  \label{eq32}
\end{equation}
where $N_{\mathrm{RF}}^{\mathrm{t}}-{{N}_{\mathrm{s}}}\le n\le {{N}_{\mathrm{s}}}$ to ensure that both ${{\mathbf{\tilde{F}}}_{\mathrm{BB1}}}$ and ${{\mathbf{\tilde{F}}}_{\mathrm{BB2}}}$ are row full-rank matrices. Different partitioning schemes can be applied in \textbf{Algorithm 1} to optimize the hybrid beamforming problem. However, as shown in Sec. VII, the spectral efficiency performance degrades as $n$ decreases, indicating that $n={{N}_{\mathrm{s}}}$ is the optimal matrix partition scheme. Although the proof of this optimality is beyond the scope of this paper, an intuitive explanation is provided in Sec. VII.

\section{Low-Complexity SVD Algorithm for Sparse Geometric Channel}
In the AREE algorithm, the optimal solution ${{\mathbf{F}}_{\mathrm{opt}}}={{\left[ \mathbf{V} \right]}_{:,1:{{N}_{\mathrm{s}}}}}$ required in (\ref{eq7}) is derived from channel's SVD $\mathbf{H}=\mathbf{U\Sigma }{{\mathbf{V}}^{H}}$. However, for mmWave XL-MIMO systems, the channel matrix $\mathbf{H}\in {{\mathbb{C}}^{{{N}_{\mathrm{r}}}\times {{N}_{\mathrm{t}}}}}$ is typically high-dimensional, resulting in significant computational complexity ($\mathcal{O}\left( {{N}_{\mathrm{t}}}N_{\mathrm{r}}^{2} \right)$ \cite{li2019tutorial}) for channel's SVD. To address this challenge, this section proposes a low-complexity SVD algorithm that exploits the sparsity inherent in the mmWave geometric channel.

\subsection{Relationship Between SVD and Geometric Channel Model}
The mmWave geometric channel (\ref{eq3}) can be expressed in the following matrix form:
\begin{equation}
  \mathbf{H}={{\mathbf{A}}_{\mathrm{r}}}{{\mathbf{H}}_{\mathrm{d}}}\mathbf{A}_{\mathrm{t}}^{H},
  \label{eq33}
\end{equation}
\begin{equation}
  {{\mathbf{A}}_{\mathrm{t}}}=\left[ {{\mathbf{a}}_{\mathrm{t}}}\left( \phi _{1,1}^{\mathrm{t}},\theta _{1,1}^{\mathrm{t}} \right),\ldots ,{{\mathbf{a}}_{\mathrm{t}}}\left( \phi _{{{N}_{\mathrm{cl}}},{{N}_{\mathrm{ray}}}}^{\mathrm{t}},\theta _{{{N}_{\mathrm{cl}}},{{N}_{\mathrm{ray}}}}^{\mathrm{t}} \right) \right],
  \label{eq34}
\end{equation}
\begin{equation}
  {{\mathbf{A}}_{\mathrm{r}}}=\left[ {{\mathbf{a}}_{\mathrm{r}}}\left( \phi _{1,1}^{\mathrm{r}},\theta _{1,1}^{\mathrm{r}} \right),\ldots ,{{\mathbf{a}}_{\mathrm{r}}}\left( \phi _{{{N}_{\mathrm{cl}}},{{N}_{\mathrm{ray}}}}^{\mathrm{r}},\theta _{{{N}_{\mathrm{cl}}},{{N}_{\mathrm{ray}}}}^{\mathrm{r}} \right) \right],
  \label{eq35}
\end{equation}
\begin{equation}
  {{\mathbf{H}}_{\mathrm{d}}}=\sqrt{\frac{{{N}_{\mathrm{t}}}{{N}_{\mathrm{r}}}}{{{N}_{\mathrm{cl}}}{{N}_{\mathrm{ray}}}}}\operatorname{diag}\left( {{\alpha }_{1,1}},\ldots ,{{\alpha }_{{{N}_{\mathrm{cl}}},{{N}_{\mathrm{ray}}}}} \right),
  \label{eq36}
\end{equation}
where ${{\mathbf{A}}_{\mathrm{t}}}\in {{\mathbb{C}}^{{{N}_{\mathrm{t}}}\times {{N}_{\mathrm{cl}}}{{N}_{\mathrm{ray}}}}}$ and ${{\mathbf{A}}_{\mathrm{r}}}\in {{\mathbb{C}}^{{{N}_{\mathrm{r}}}\times {{N}_{\mathrm{cl}}}{{N}_{\mathrm{ray}}}}}$ are steering matrices that consist of all channel steering vectors for the transmitter and receiver, respectively. ${{\mathbf{H}}_{\mathrm{d}}}\in {{\mathbb{C}}^{{{N}_{\mathrm{cl}}}{{N}_{\mathrm{ray}}}\times {{N}_{\mathrm{cl}}}{{N}_{\mathrm{ray}}}}}$ is a diagonal matrix composed of complex channel gains from all propagation paths. Before introducing the proposed algorithm, we make the following assumption:

\textit{Assumption 1:} Under the assumption that $\mathbf{H}$ is perfectly known, we assume that ${{\mathbf{H}}_{\mathrm{d}}}$, ${{\mathbf{A}}_{\mathrm{r}}}$ and ${{\mathbf{A}}_{\mathrm{t}}}$ are known as well. 

This assumption is reasonable because in mmWave XL-MIMO, $\mathbf{H}$ is typically derived after obtaining ${{\mathbf{H}}_{\mathrm{d}}}$, ${{\mathbf{A}}_{\mathrm{r}}}$ and ${{\mathbf{A}}_{\mathrm{t}}}$. Specifically, because of the high dimensionality of $\mathbf{H}$, mmWave XL-MIMO channel estimation commonly employs sparse channel estimation algorithms \cite{alkhateeb2014channel,heath2016overview,lee2016channel,rodriguez2018frequency}, such as OMP-based or Bayesian compressive sensing methods, to first estimate ${{\mathbf{H}}_{\mathrm{d}}}$, ${{\mathbf{A}}_{\mathrm{r}}}$ and ${{\mathbf{A}}_{\mathrm{t}}}$. The overall channel $\mathbf{H}$ is then reconstructed by multiplying these matrices as described in (\ref{eq33}).

The channel SVD can be expressed as $\mathbf{H}=\mathbf{U\Sigma }{{\mathbf{V}}^{H}}$, where $\mathbf{\Sigma }\in {{\mathbb{C}}^{K\times K}}$, $\mathbf{U}\in {{\mathbb{C}}^{{{N}_{\mathrm{r}}}\times K}}$ and $\mathbf{V}\in {{\mathbb{C}}^{{{N}_{\mathrm{t}}}\times K}}$ represent the singular value matrix (real diagonal matrix), and the left and right singular matrix (semi-unitary matrix), respectively, where $K=\operatorname{rank}\left( \mathbf{H} \right)$. Despite similar in structure, the geometric channel model in (\ref{eq33}) and the channel SVD are different: ${{\mathbf{H}}_{\mathrm{d}}}$ is not a real matrix, ${{\mathbf{A}}_{\mathrm{r}}}$ and ${{\mathbf{A}}_{\mathrm{t}}}$ are not semi-unitary matrices. This indicates that neither ${{\mathbf{A}}_{\mathrm{r}}}$ nor ${{\mathbf{A}}_{\mathrm{t}}}$ is a column-orthogonal matrix, meaning that there may be correlations between ${{\mathbf{a}}_{\mathrm{r}}}\left( \phi _{i,{{l}_{1}}}^{\mathrm{r}},\theta _{i,{{l}_{1}}}^{\mathrm{r}} \right)$ and ${{\mathbf{a}}_{\mathrm{r}}}\left( \phi _{i,{{l}_{2}}}^{\mathrm{r}},\theta _{i,{{l}_{2}}}^{\mathrm{r}} \right)$, or ${{\mathbf{a}}_{\mathrm{t}}}\left( \phi _{i,{{l}_{1}}}^{\mathrm{t}},\theta _{i,{{l}_{1}}}^{\mathrm{t}} \right)$ and ${{\mathbf{a}}_{\mathrm{t}}}\left( \phi _{i,{{l}_{2}}}^{\mathrm{t}},\theta _{i,{{l}_{2}}}^{\mathrm{t}} \right)$ when ${{l}_{1}}\ne {{l}_{2}}$. This correlation results from insufficient angle resolution of the antenna array: different propagation paths within the same cluster may have similar angles, and when the antenna array cannot fully distinguish these angular differences, steering vectors of different propagation paths will not be orthogonal, resulting in ${{\mathbf{A}}_{\mathrm{r}}}$ and ${{\mathbf{A}}_{\mathrm{t}}}$ typically not being column-orthogonal matrices.

\begin{algorithm}[t]
  \caption{Geometric channel SVD (GC-SVD) algorithm}\label{algorithm3}
  \begin{algorithmic}[1]
      \Require{${{\mathbf{A}}_{\mathrm{r}}}$, ${{\mathbf{H}}_{\mathrm{d}}}$, ${{\mathbf{A}}_{\mathrm{t}}}$}
      \State
      Perform SVD: ${{\mathbf{A}}_{\mathrm{r}}}=\mathbf{U}_{\mathrm{r}}\mathbf{\Sigma }_{\mathrm{R}}^{{1}/{2}\;}\mathbf{Q}^{H}$;
      \State
      Perform SVD: ${{\mathbf{A}}_{\mathrm{t}}}=\mathbf{U}_{\mathrm{t}}\mathbf{\Sigma }_{\mathrm{T}}^{{1}/{2}\;}\mathbf{P}^{H}$;
      \State
      $\mathbf{\tilde{Q}}=\mathbf{Q\Sigma }_{\mathrm{R}}^{-{1}/{2}\;}$, $\mathbf{\tilde{P}}=\mathbf{P\Sigma }_{\mathrm{T}}^{-{1}/{2}\;}$, ${{\mathbf{\tilde{H}}}_{\mathrm{d}}}={{\mathbf{\Sigma }}_{\mathrm{R}}}{{\mathbf{\tilde{Q}}}^{H}}{{\mathbf{H}}_{\mathrm{d}}}\mathbf{\tilde{P}}{{\mathbf{\Sigma }}_{\mathrm{T}}}$;
      \State
      Perform SVD: ${{\mathbf{\tilde{H}}}_{\mathrm{d}}}=\mathbf{\tilde{U}\Sigma }{{\mathbf{\tilde{V}}}^{H}}$;
      \State
      $\mathbf{U}={{\mathbf{A}}_{\mathrm{r}}}\mathbf{\tilde{Q}\tilde{U}}$, $\mathbf{V}={{\mathbf{A}}_{\mathrm{t}}}\mathbf{\tilde{P}\tilde{V}}$;
      \Ensure
      $\mathbf{U}$, $\mathbf{\Sigma }$, $\mathbf{V}$
  \end{algorithmic}
\end{algorithm}

\subsection{From a High-Dimensional Sparse Channel to a Low-Dimensional Non-Sparse Channel}

\textit{Lemma 1:} Let $R=\operatorname{rank}\left( {{\mathbf{A}}_{\mathrm{r}}} \right)$ and $T=\operatorname{rank}\left( {{\mathbf{A}}_{\mathrm{t}}} \right)$, where $R\le {{N}_{\mathrm{cl}}}{{N}_{\mathrm{ray}}}$ and $T\le {{N}_{\mathrm{cl}}}{{N}_{\mathrm{ray}}}$. There exist $\mathbf{\tilde{Q}}\in {{\mathbb{C}}^{{{N}_{\mathrm{cl}}}{{N}_{\mathrm{ray}}}\times R}}$ and $\mathbf{\tilde{P}}\in {{\mathbb{C}}^{{{N}_{\mathrm{cl}}}{{N}_{\mathrm{ray}}}\times T}}$ to make ${{\mathbf{\tilde{A}}}_{\mathrm{r}}}={{\mathbf{A}}_{\mathrm{r}}}\mathbf{\tilde{Q}}\in {{\mathbb{C}}^{{{N}_{\mathrm{r}}}\times R}}$ and ${{\mathbf{\tilde{A}}}_{\mathrm{t}}}={{\mathbf{A}}_{\mathrm{t}}}\mathbf{\tilde{P}}\in {{\mathbb{C}}^{{{N}_{\mathrm{t}}}\times T}}$ be semi-unitary matrices, respectively.

\textit{Proof:} Given the SVD of ${{\mathbf{A}}_{\mathrm{r}}}$ as ${{\mathbf{A}}_{\mathrm{r}}}={{\mathbf{U}}_{\mathrm{r}}}\mathbf{\Sigma }_{\mathrm{R}}^{{1}/{2}\;}{{\mathbf{Q}}^{H}}$, where ${{\mathbf{\Sigma }}_{\mathrm{R}}}\in {{\mathbb{C}}^{R\times R}}$ is a real diagonal matrix and $\mathbf{Q}\in {{\mathbb{C}}^{{{N}_{\mathrm{cl}}}{{N}_{\mathrm{ray}}}\times R}}$ is a semi-unitary matrix, define $\mathbf{\tilde{Q}}=\mathbf{Q\Sigma }_{\mathrm{R}}^{-{1}/{2}\;}$. Then we have ${{\left( {{\mathbf{A}}_{\mathrm{r}}}\mathbf{\tilde{Q}} \right)}^{H}}{{\mathbf{A}}_{\mathrm{r}}}\mathbf{\tilde{Q}}={{\mathbf{I}}_{\mathrm{R}}}$, which implies that ${{\mathbf{\tilde{A}}}_{\mathrm{r}}}={{\mathbf{A}}_{\mathrm{r}}}\mathbf{\tilde{Q}}$ is a semi-unitary matrix. Similarly, from the SVD of ${{\mathbf{A}}_{\mathrm{t}}}$ as ${{\mathbf{A}}_{\mathrm{t}}}={{\mathbf{U}}_{\mathrm{t}}}\mathbf{\Sigma }_{\mathrm{T}}^{{1}/{2}\;}{{\mathbf{P}}^{H}}$, where ${{\mathbf{\Sigma }}_{\mathrm{T}}}\in {{\mathbb{C}}^{T\times T}}$ is a real diagonal matrix and $\mathbf{P}\in {{\mathbb{C}}^{{{N}_{\mathrm{cl}}}{{N}_{\mathrm{ray}}}\times T}}$ is a semi-unitary matrix, we define $\mathbf{\tilde{P}}=\mathbf{P\Sigma }_{\mathrm{T}}^{-{1}/{2}\;}$ and then we have ${{\left( {{\mathbf{A}}_{\mathrm{t}}}\mathbf{\tilde{P}} \right)}^{H}}{{\mathbf{A}}_{\mathrm{t}}}\mathbf{\tilde{P}}={{\mathbf{I}}_{\mathrm{T}}}$, which implies that ${{\mathbf{\tilde{A}}}_{\mathrm{t}}}={{\mathbf{A}}_{\mathrm{t}}}\mathbf{\tilde{P}}$ is a semi-unitary matrix.\hfill$\blacksquare$

Consequently, the geometric channel in (\ref{eq33}) can be further expressed as
\begin{equation}
  \begin{aligned}
  & \mathbf{H}={{\mathbf{U}}_{\mathrm{r}}}\mathbf{\Sigma }_{\mathrm{R}}^{{1}/{2}\;}{{\mathbf{Q}}^{H}}\mathbf{Q}{{\mathbf{Q}}^{H}}{{\mathbf{H}}_{\mathrm{d}}}{{\left( {{\mathbf{U}}_{\mathrm{t}}}\mathbf{\Sigma }_{\mathrm{T}}^{{1}/{2}\;}{{\mathbf{P}}^{H}}\mathbf{P}{{\mathbf{P}}^{H}} \right)}^{H}} \\ 
 & \quad ={{\mathbf{A}}_{\mathrm{r}}}\mathbf{Q}\left( {{\mathbf{Q}}^{H}}{{\mathbf{H}}_{\mathrm{d}}}\mathbf{P} \right){{\left( {{\mathbf{A}}_{\mathrm{t}}}\mathbf{P} \right)}^{H}} \\ 
 & \quad ={{\mathbf{A}}_{\mathrm{r}}}\mathbf{Q\Sigma }_{\mathrm{R}}^{-\!{1}\!/{2}\;}\!\!\!\left( \!{{\mathbf{\Sigma }}_{\mathrm{R}}}\mathbf{\Sigma }_{\mathrm{R}}^{-\!{1}\!/{2}\;}\!\!{{\mathbf{Q}}^{H}}{{\mathbf{H}}_{\mathrm{d}}}\mathbf{P\Sigma }_{\mathrm{T}}^{-\!{1}\!/{2}\;}\!{{\mathbf{\Sigma }}_{\mathrm{T}}}\! \right)\!{\!{\left( {{\mathbf{A}}_{\mathrm{t}}}\mathbf{P\Sigma }_{\mathrm{T}}^{-\!{1}\!/{2}\;}\!\! \right)}^{\!\!H}} \\ 
 & \quad =\left( {{\mathbf{A}}_{\mathrm{r}}}\mathbf{\tilde{Q}} \right)\left( {{\mathbf{\Sigma }}_{\mathrm{R}}}{{{\mathbf{\tilde{Q}}}}^{H}}{{\mathbf{H}}_{\mathrm{d}}}\mathbf{\tilde{P}}{{\mathbf{\Sigma }}_{\mathrm{T}}} \right){{\left( {{\mathbf{A}}_{\mathrm{t}}}\mathbf{\tilde{P}} \right)}^{H}} \\ 
 & \quad ={{{\mathbf{\tilde{A}}}}_{\mathrm{r}}}{{{\mathbf{\tilde{H}}}}_{\mathrm{d}}}\mathbf{\tilde{A}}_{\mathrm{t}}^{H}\ , \\ 
\end{aligned}
\label{eq37}
\end{equation}
where ${{\mathbf{\tilde{H}}}_{\mathrm{d}}}={{\mathbf{\Sigma }}_{\mathrm{R}}}{{\mathbf{\tilde{Q}}}^{H}}{{\mathbf{H}}_{\mathrm{d}}}\mathbf{\tilde{P}}{{\mathbf{\Sigma }}_{\mathrm{T}}}$. As a result, ${{\mathbf{A}}_{\mathrm{r}}}$ and ${{\mathbf{A}}_{\mathrm{t}}}$ are transformed into semi-unitary matrices ${{\mathbf{\tilde{A}}}_{\mathrm{r}}}$ and ${{\mathbf{\tilde{A}}}_{\mathrm{t}}}$, while ${{\mathbf{H}}_{\mathrm{d}}}$ is reduced to an equivalent ``low-dimensional non-sparse" channel ${{\mathbf{\tilde{H}}}_{\mathrm{d}}}\in {{\mathbb{C}}^{R\times T}}$. Consequently, the high-dimensional SVD problem of $\mathbf{H}$ is converted into a low-dimensional SVD problem of ${{\mathbf{\tilde{H}}}_{\mathrm{d}}}$.

\subsection{Perform SVD on a Low-Dimensional Non-Sparse Channel}
The SVD of the low-dimensional non-sparse channel ${{\mathbf{\tilde{H}}}_{\mathrm{d}}}$ can be represented as ${{\mathbf{\tilde{H}}}_{\mathrm{d}}}=\mathbf{\tilde{U}\Sigma }{{\mathbf{\tilde{V}}}^{H}}$, where $\mathbf{\Sigma }\in {{\mathbb{C}}^{K\times K}}$, $\mathbf{\tilde{U}}\in {{\mathbb{C}}^{R\times K}}$ and $\mathbf{\tilde{V}}\in {{\mathbb{C}}^{T\times K}}$. Substituting this SVD into (\ref{eq37}), we finally obtain the SVD of $\mathbf{H}$ as 
\begin{equation}
  \mathbf{H}={{\mathbf{\tilde{A}}}_{\mathrm{r}}}\mathbf{\tilde{U}\Sigma }{{\mathbf{\tilde{V}}}^{H}}\mathbf{\tilde{A}}_{\mathrm{t}}^{H}=\mathbf{U\Sigma }{{\mathbf{V}}^{H}},
  \label{eq38}
\end{equation}
\begin{equation}
  \mathbf{U}={{\mathbf{\tilde{A}}}_{\mathrm{r}}}\mathbf{\tilde{U}}={{\mathbf{A}}_{\mathrm{r}}}\mathbf{\tilde{Q}\tilde{U}},
  \label{eq39}
\end{equation}
\begin{equation}
  \mathbf{V}={{\mathbf{\tilde{A}}}_{\mathrm{t}}}\mathbf{\tilde{V}}={{\mathbf{A}}_{\mathrm{t}}}\mathbf{\tilde{P}\tilde{V}},
  \label{eq40}
\end{equation}
\begin{equation}
  \mathbf{\Sigma }={{\mathbf{\tilde{U}}}^{H}}{{\mathbf{\tilde{H}}}_{\mathrm{d}}}\mathbf{\tilde{V}}={{\left( \mathbf{\tilde{Q}}{{\mathbf{\Sigma }}_{\mathrm{R}}}\mathbf{\tilde{U}} \right)}^{H}}{{\mathbf{H}}_{\mathrm{d}}}\mathbf{\tilde{P}}{{\mathbf{\Sigma }}_{\mathrm{T}}}\mathbf{\tilde{V}},
  \label{eq41}
\end{equation}
where $\mathbf{\Sigma }\in {{\mathbb{C}}^{K\times K}}$, $\mathbf{U}\in {{\mathbb{C}}^{{{N}_{\mathrm{r}}}\times K}}$ and $\mathbf{V}\in {{\mathbb{C}}^{{{N}_{\mathrm{t}}}\times K}}$ represent the singular value matrix, and the left and right singular matrices of $\mathbf{H}$, respectively. 

The main steps of the proposed SVD algorithm for sparse geometric channel (GC-SVD) are summarized in  \textbf{Algorithm~2}. By decomposing the high-dimensional SVD of $\mathbf{H}$ into three low-dimensional SVDs, the overall complexity is reduced to $\mathcal{O}\left( \left( {{N}_{\mathrm{t}}}+{{N}_{\mathrm{r}}} \right){{\left( {{N}_{\mathrm{cl}}}{{N}_{\mathrm{ray}}} \right)}^{2}} \right)$, which is lower than that of the traditional SVD algorithm given by $\mathcal{O}\left( {{N}_{\mathrm{t}}}N_{\mathrm{r}}^{2} \right)$ \cite{li2019tutorial}. In XL-MIMO communication systems, such as for vehicular networks or scenarios with a limited number of line-of-sight propagation paths, it is reasonable that ${{N}_{\mathrm{r}}}>{{N}_{\mathrm{cl}}}{{N}_{\mathrm{ray}}}$ and the proposed GC-SVD algorithm effectively reduces complexity. In addition, it establishes a linear relationship between the singular matrices $\mathbf{U}$, $\mathbf{V}$ and the steering matrices ${{\mathbf{A}}_{\mathrm{r}}}$, ${{\mathbf{A}}_{\mathrm{t}}}$ (as illustrated in (\ref{eq39}) and (\ref{eq40})), which can be further used to simplify the initialization method proposed later. 

\section{Initialization Methods for Hybrid Beamforming}
AREE algorithm approximates the suboptimal solution through iterative optimizations, as detailed in Sec.~VII. However, iterative algorithms often incur high computational complexity, particularly in XL-MIMO systems. A well-chosen initialization can significantly accelerate convergence and reduce complexity. In this context, low-complexity non-iterative hybrid beamforming algorithms can offer effective initial values for iterative methods.

In this section, we first propose a low-complexity non-iterative hybrid beamforming scheme, whose complexity is further reduced on the basis of the GC-SVD algorithm. As presented in Sec. VI and Sec. VII, the proposed methods outperform existing OMP-based algorithms with a lower complexity, providing effective initial values for AREE algorithm.

\begin{algorithm}[t]
  \caption{PE-OMP hybrid beamforming algorithm}\label{algorithm4}
  \begin{algorithmic}[1]
      \Require{${{\mathbf{F}}_{\mathrm{opt}}}$, ${{\mathbf{A}}_{\mathrm{t}}}$}
      \State
      \textbf{Initialization:} ${{\mathbf{F}}_{\mathrm{\!res}}}\!=\!{{\mathbf{F}}_{\mathrm{\!opt}}}$, $\mathbf{A}_{\mathrm{t}}^{\max }$ is an empty matrix, $i\!\!=\!\!1$;
      \For{$i\le N_{\mathrm{RF}}^{\mathrm{t}}-{{N}_{\mathrm{s}}}$}
      \State
      $\mathbf{\Psi }=\mathbf{A}_{\mathrm{t}}^{H}{{\mathbf{F}}_{\mathrm{res}}}$;
      \State
      $k=\underset{l=1,\ldots ,{{N}_{\mathrm{cl}}}{{N}_{\mathrm{ray}}}}{\mathop{\arg \max }}\,{{\left\| {{\left[ \mathbf{\Psi } \right]}_{l,:}} \right\|}_{2}}$;
      \State
      $\mathbf{A}_{\mathrm{t}}^{\max }=\left[ \mathbf{A}_{\mathrm{t}}^{\max },\ {{\left[ {{\mathbf{A}}_{\mathrm{t}}} \right]}_{:,k}} \right]$;
      \State
      ${{\mathbf{F}}_{\mathrm{BB}}}={{\left( \mathbf{A}_{\mathrm{t}}^{\max } \right)}^{\dagger }}{{\mathbf{F}}_{\mathrm{opt}}}$;
      \State
      ${{\mathbf{F}}_{\mathrm{res}}}={{\mathbf{F}}_{\mathrm{opt}}}-\mathbf{A}_{\mathrm{t}}^{\max }{{\mathbf{F}}_{\mathrm{BB}}}$;
      \State
      $i=i+1$;
      \EndFor
      \State
      \textbf{end for}
      \State
      ${{\mathbf{F}}_{\mathrm{RF}}^{\mathrm{initial}}}=\frac{1}{\sqrt{{{N}_{\mathrm{t}}}}}\exp \left( \mathrm{j}\cdot \arg \left( \left[ {{\mathbf{F}}_{\mathrm{res}}}\ ,\ \mathbf{A}_{\mathrm{t}}^{\max } \right] \right) \right)$;
      \State
      ${{\mathbf{F}}_{\mathrm{BB}}}=\mathbf{F}_{\mathrm{RF}}^{\dagger }{{\mathbf{F}}_{\mathrm{opt}}},\quad {{\mathbf{F}}_{\mathrm{BB}}^{\mathrm{initial}}}=\frac{\sqrt{{{N}_{\mathrm{s}}}}}{{{\left\| {{\mathbf{F}}_{\mathrm{RF}}}{{\mathbf{F}}_{\mathrm{BB}}} \right\|}_{F}}}{{\mathbf{F}}_{\mathrm{BB}}}$;
      \Ensure
      ${{\mathbf{F}}_{\mathrm{RF}}^{\mathrm{initial}}}$, ${{\mathbf{F}}_{\mathrm{BB}}^{\mathrm{initial}}}$
  \end{algorithmic}
\end{algorithm}

\subsection{PE-OMP Algorithm}
We first introduce the phase extraction with OMP (PE-OMP) algorithm, as described in the pseudo-code provided in \textbf{Algorithm 3}. It first selects $N_{\mathrm{RF}}^{\mathrm{t}}-{{N}_{\mathrm{s}}}$ steering vectors from the channel steering matrix ${{\mathbf{A}}_{\mathrm{t}}}$ that have the largest projections onto ${{\mathbf{F}}_{\mathrm{opt}}}$, and these vectors are then combined to form $\mathbf{A}_{\mathrm{t}}^{\max }$. After each selection, ${{\left( \mathbf{A}_{\mathrm{t}}^{\max } \right)}^{\dagger }}{{\mathbf{F}}_{\mathrm{BB}}}$ is subtracted from ${{\mathbf{F}}_{\mathrm{res}}}$ to mitigate the impact of the selected steering vectors on the future selections. Ultimately, $\mathbf{A}_{\mathrm{t}}^{\max }\in {{\mathbb{C}}^{{{N}_{\mathrm{t}}}\times \left( N_{\mathrm{RF}}^{\mathrm{t}}-{{N}_{\mathrm{s}}} \right)}}$ consists of the selected steering vectors, while ${{\mathbf{F}}_{\mathrm{res}}}\in {{\mathbb{C}}^{{{N}_{\mathrm{t}}}\times {{N}_{\mathrm{s}}}}}$ comprises the remaining unselected steering vectors. Then, both $\mathbf{A}_{\mathrm{t}}^{\max }$ and ${{\mathbf{F}}_{\mathrm{res}}}$ are constrained to have constant modulus values to construct the analog precoder ${{\mathbf{F}}_{\mathrm{RF}}}\in {{\mathbb{C}}^{{{N}_{\mathrm{t}}}\times N_{\mathrm{RF}}^{\mathrm{t}}}}$.

Compared to existing OMP-based algorithms \cite{el2014spatially,lee2014hybrid,chen2016compressive}, the PE-OMP algorithm offers two advantages. First, we 
select only ${N_{\mathrm{RF}}^{\mathrm{t}}-{{N}_{\mathrm{s}}}}$ of the strongest steering vectors from $\mathbf{A}_{\mathrm{t}}$ to construct 
$\mathbf{A}_{\mathrm{t}}^{\max }$, whereas OMP selects ${N}_{\mathrm{RF}}^{\mathrm{t}}$ vectors, so the PE-OMP achieves lower computational complexity. Second, the remaining steering vectors in $\mathbf{A}_{\mathrm{t}}$ construct ${{\mathbf{F}}_{\mathrm{res}}}$ in PE-OMP, while they are discarded in OMP. Although ${{\mathbf{F}}_{\mathrm{res}}}$ experiences distortions due to constant modulus operation, it retains all remaining vectors which are also useful in the reconstruction of precoders, where the energy of ${{\mathbf{F}}_{\mathrm{res}}}$ may even exceed that of $\mathbf{A}_{\mathrm{t}}^{\max }$ when $N_{\mathrm{RF}}^{\mathrm{t}}$ is small, thereby further improving beamforming performance. As a result, the proposed PE-OMP algorithm outperforms existing OMP-based algorithms in spectral efficiency while offering a reduced complexity.

\subsection{PE-SMD Algorithm}
Based on the linear relationship between the channel's singular vectors and the steering vectors established by the GC-SVD method in (\ref{eq40}), we propose the phase extraction with singular matrix decomposition (PE-SMD) algorithm to further reduce computational complexity of the PE-OMP. As detailed in the pseudo-code of \textbf{Algorithm 4}, PE-SMD simplifies PE-OMP in the following three key aspects, incurring only a slight performance degradation.

\subsubsection{Omit Projection Operations}
PE-SMD omits the projection operation $\mathbf{\Psi }=\mathbf{A}_{\mathrm{t}}^{H}{{\mathbf{F}}_{\mathrm{res}}}$, as the relationship
\begin{equation}
 {{\mathbf{F}}_{\mathrm{opt}}}={{\left[ \mathbf{V} \right]}_{:,1:{{N}_{\mathrm{s}}}}}={{\mathbf{A}}_{\mathrm{t}}}{{\left[ \mathbf{\tilde{P}\tilde{V}} \right]}_{:,1:{{N}_{\mathrm{s}}}}},
 \label{eq42}
\end{equation}
 which is described in (\ref{eq6}) and (\ref{eq40}), indicates that {\small ${{\left[ \mathbf{\tilde{P}\tilde{V}} \right]}_{:,1:{{N}_{\mathrm{s}}}}}$} can approximately replace $\mathbf{\Psi }$ as the projection of $\mathbf{A}_{\mathrm{t}}$ onto ${{\mathbf{F}}_{\mathrm{opt}}}$. 

\subsubsection{Select Steering Vectors Simultaneously}
PE-SMD simultaneously selects $N_{\mathrm{RF}}^{\mathrm{t}}-{{N}_{\mathrm{s}}}$ vectors in $\mathbf{A}_{\mathrm{t}}$ which have the largest projections onto $\mathbf{F}_{\mathrm{opt}}$ to construct ${\mathbf{A}_{\mathrm{t}}^{\mathrm{max}}}={{\left[ {{\mathbf{A}}_{\mathrm{t}}} \right]}_{:,\mathcal{F}}}$, where $\mathcal{F}$ is defined as
\begin{equation}
  \mathcal{F}=\left\{ {{l}_{1}},\ldots ,{{l}_{{{{N}_{\mathrm{\!R\!F\!}}^{\mathrm{t}}-{{N}_{\mathrm{\!s}}}}}}} \right\}=\underset{l=1,\ldots ,{{N}_{\mathrm{cl}}}{{N}_{\mathrm{ray}}}}{\mathop{\arg \max }}\,{{\left\| {{\left[ \mathbf{\tilde{P}\tilde{V}} \right]}_{l,1:{{N}_{\mathrm{s}}}}} \right\|}_{\mathrm{2}}}.
\label{eq43}
\end{equation}

\subsubsection{Replace Pseudo-Inverse with Conjugate Transpose}
PE-SMD utilizes ${{\left( {{\left[ {{\mathbf{A}}_{\mathrm{t}}} \right]}_{:,\mathcal{F}}} \right)}^{H}}$ instead of ${{\left( {{\left[ {{\mathbf{A}}_{\mathrm{t}}} \right]}_{:,\mathcal{F}}} \right)}^{\dagger }}$ to compute ${{\mathbf{P}}_{\mathrm{pjt}}}$, which is the primary reason for performance loss compared to PE-OMP. However, since ${{\mathbf{A}}_{\mathrm{t}}}$ is approximately a semi-unitary matrix (because the columns of ${{\mathbf{A}}_{\mathrm{t}}}\in {{\mathbb{C}}^{{{N}_{\mathrm{t}}}\times {{N}_{\mathrm{cl}}}{{N}_{\mathrm{ray}}}}}$ exhibit minimal correlation for large values of ${{N}_{\mathrm{t}}}$ ), the performance loss remains slight while complexity being significantly reduced.

\begin{figure*}[t]
  \centering
  \subfloat[$\mathbf{H}_{\mathrm{eq}}^{\mathrm{opt}}$ of the fully-digital algorithm]
  {\label{fig3a}\includegraphics[width=0.28\textwidth]{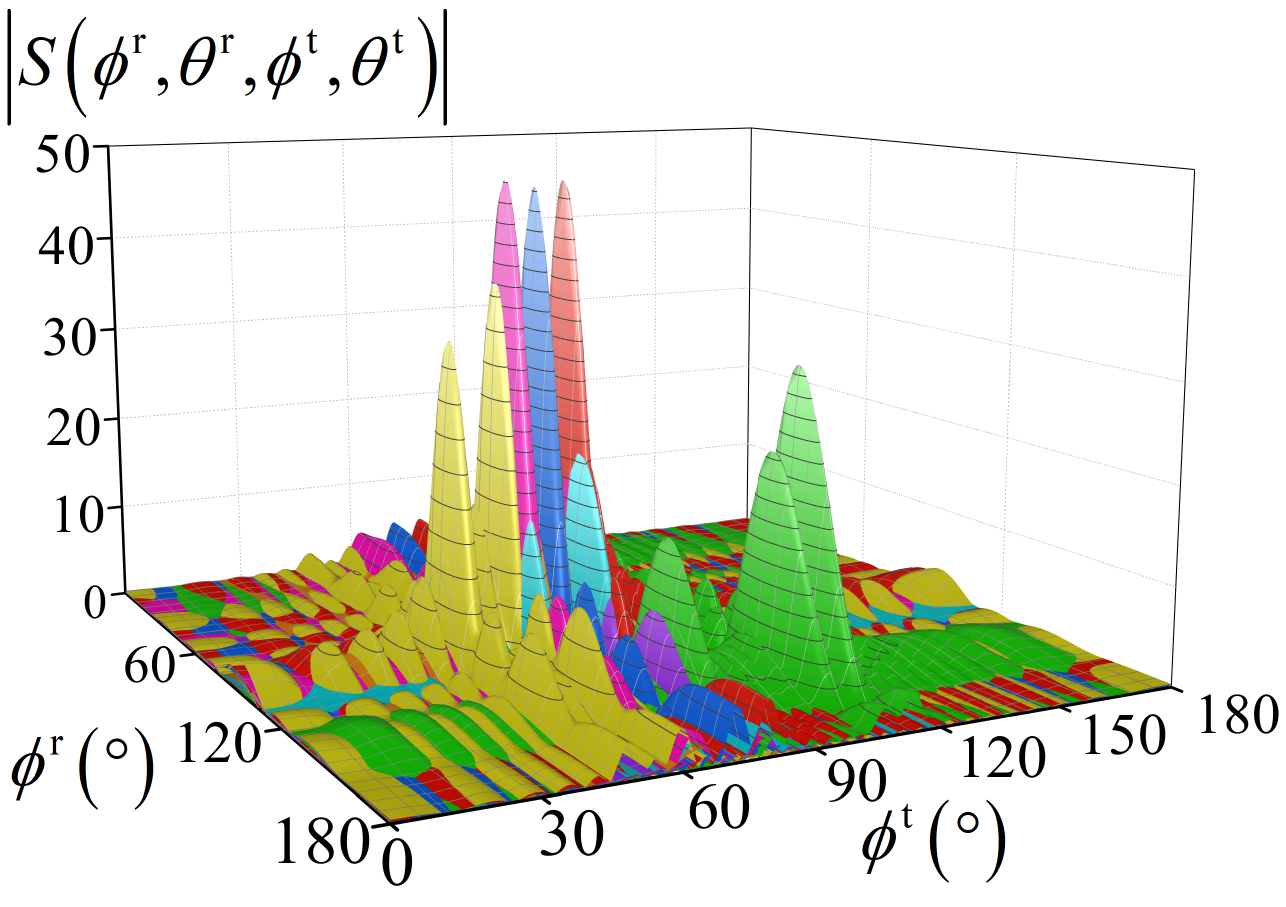}}
  \hspace{0.5cm} 
  \subfloat[$\mathbf{H}_{\mathrm{eq}}$ of the OMP algorithm]
  {\label{fig3b}\includegraphics[width=0.28\textwidth]{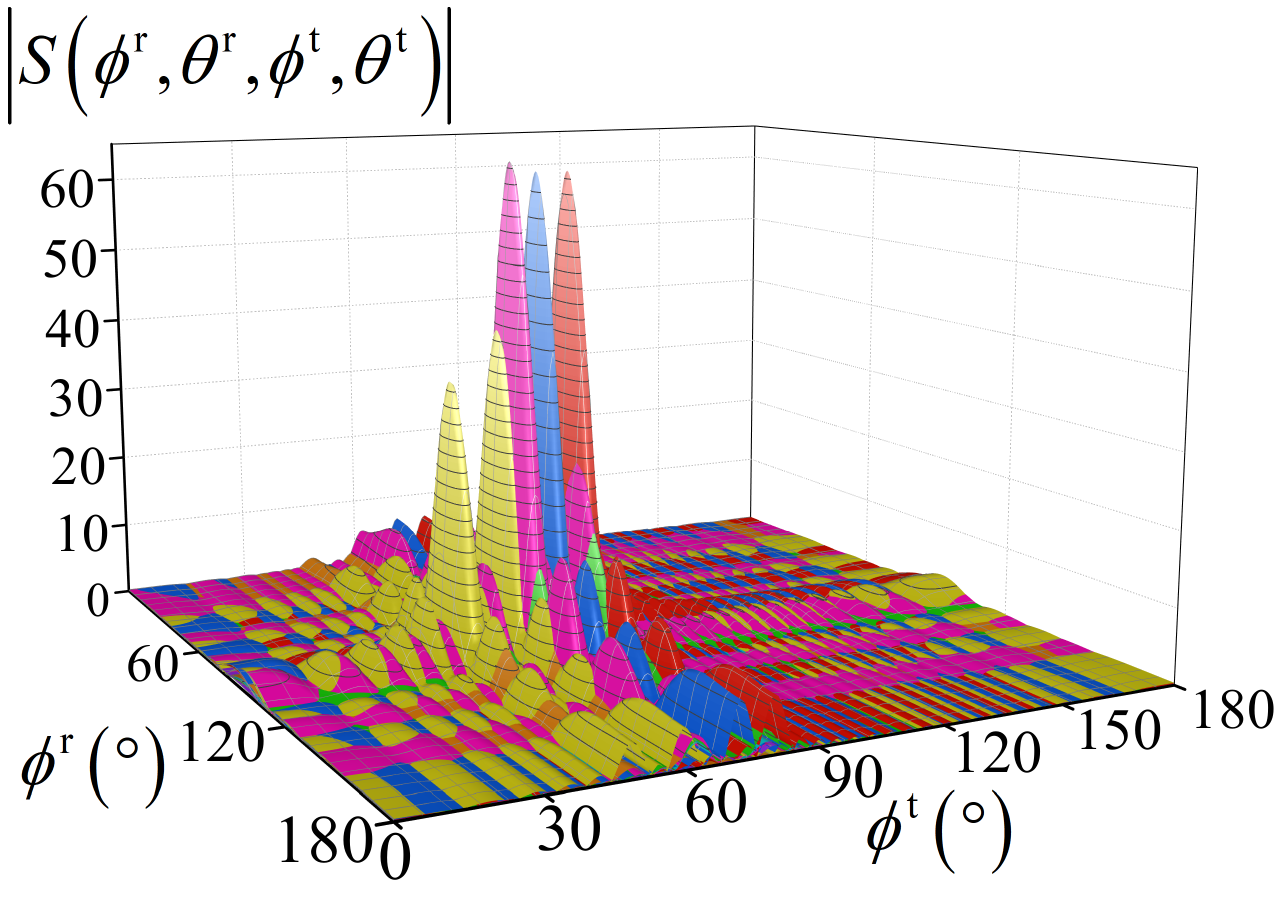}}
  \hspace{0.5cm} 
  \subfloat[$\mathbf{H}_{\mathrm{eq}}$ of the PE-OMP algorithm]
  {\label{fig3c}\includegraphics[width=0.28\textwidth]{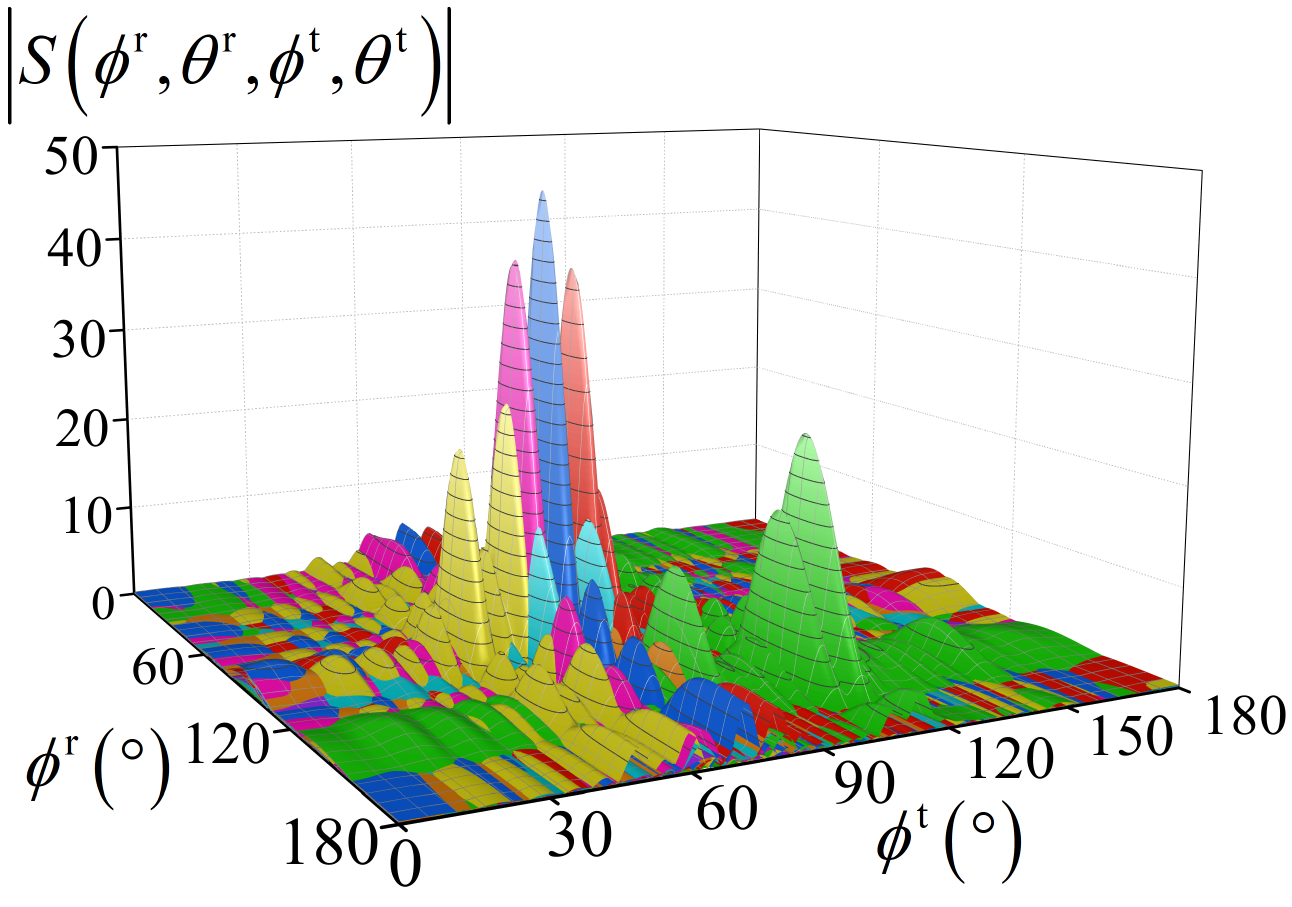}}

  \subfloat[$\mathbf{H}_{\mathrm{eq}}$ of the channel singular vector]
  {\label{fig3d}\includegraphics[width=0.28\textwidth]{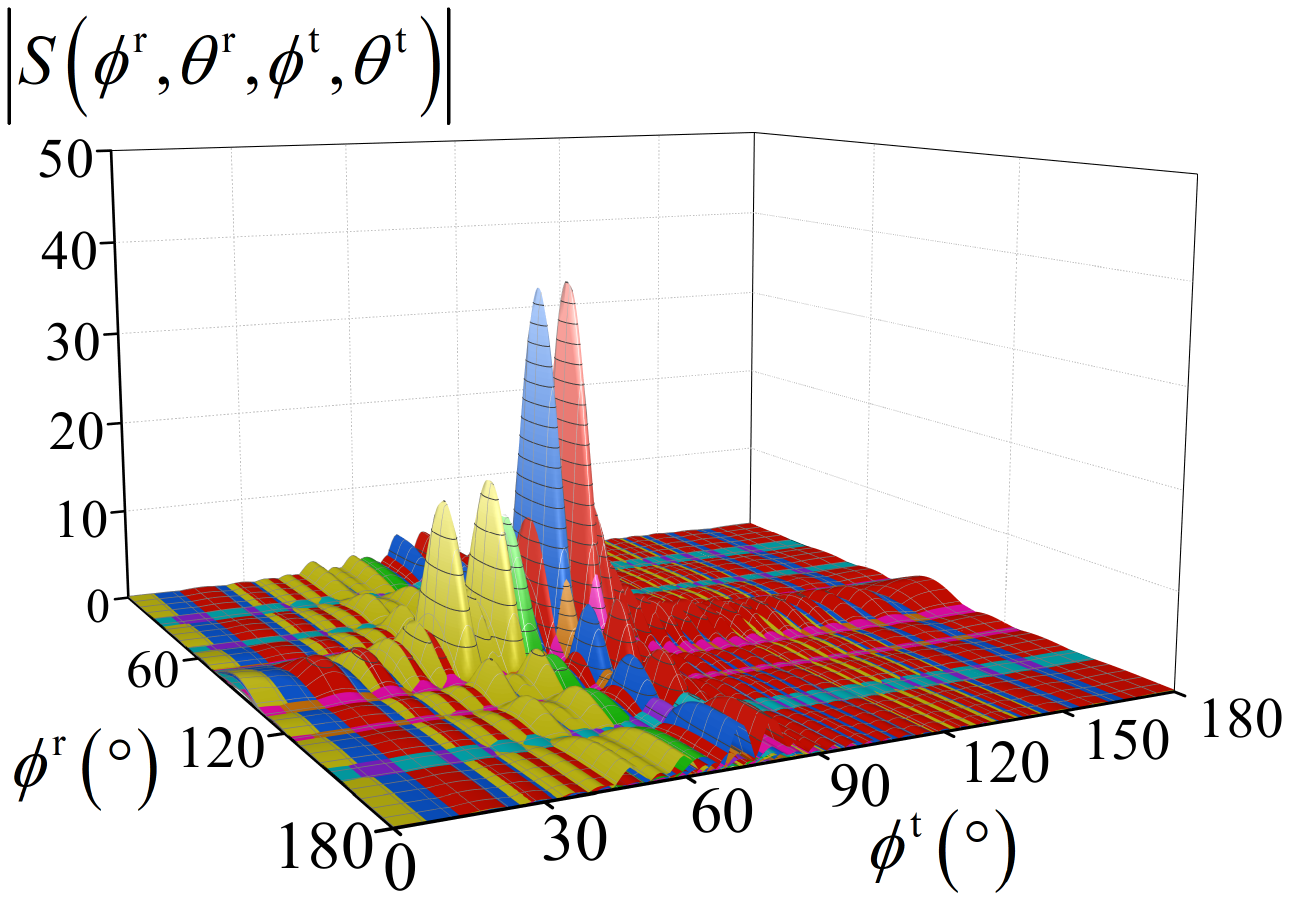}}
  \hspace{0.5cm} 
  \subfloat[$\mathbf{H}_{\mathrm{gap}}$ of the OMP algorithm]
  {\label{fig3e}\includegraphics[width=0.28\textwidth]{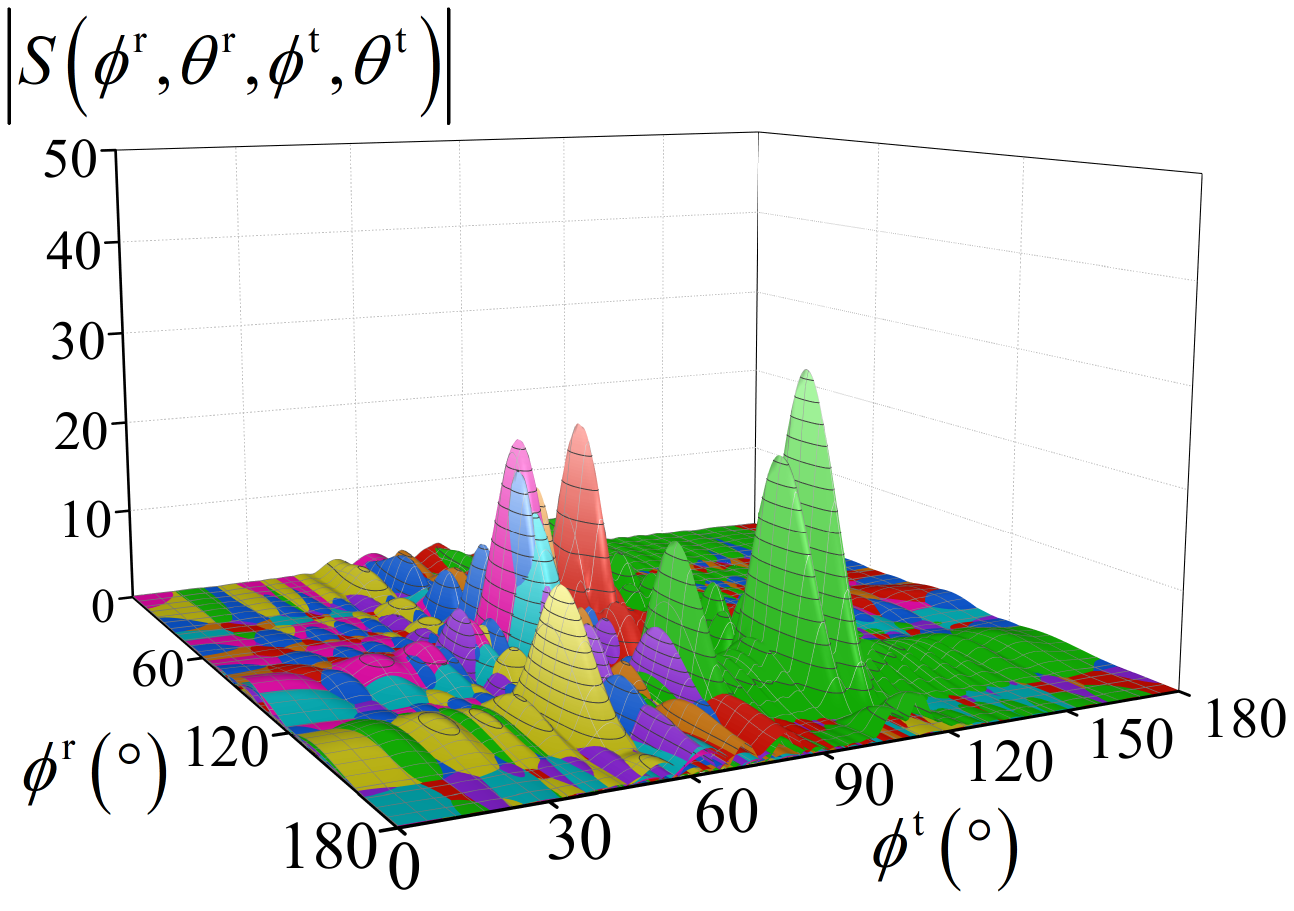}}
  \hspace{0.5cm} 
  \subfloat[$\mathbf{H}_{\mathrm{gap}}$ of the PE-OMP algorithm]
  {\label{fig3f}\includegraphics[width=0.28\textwidth]{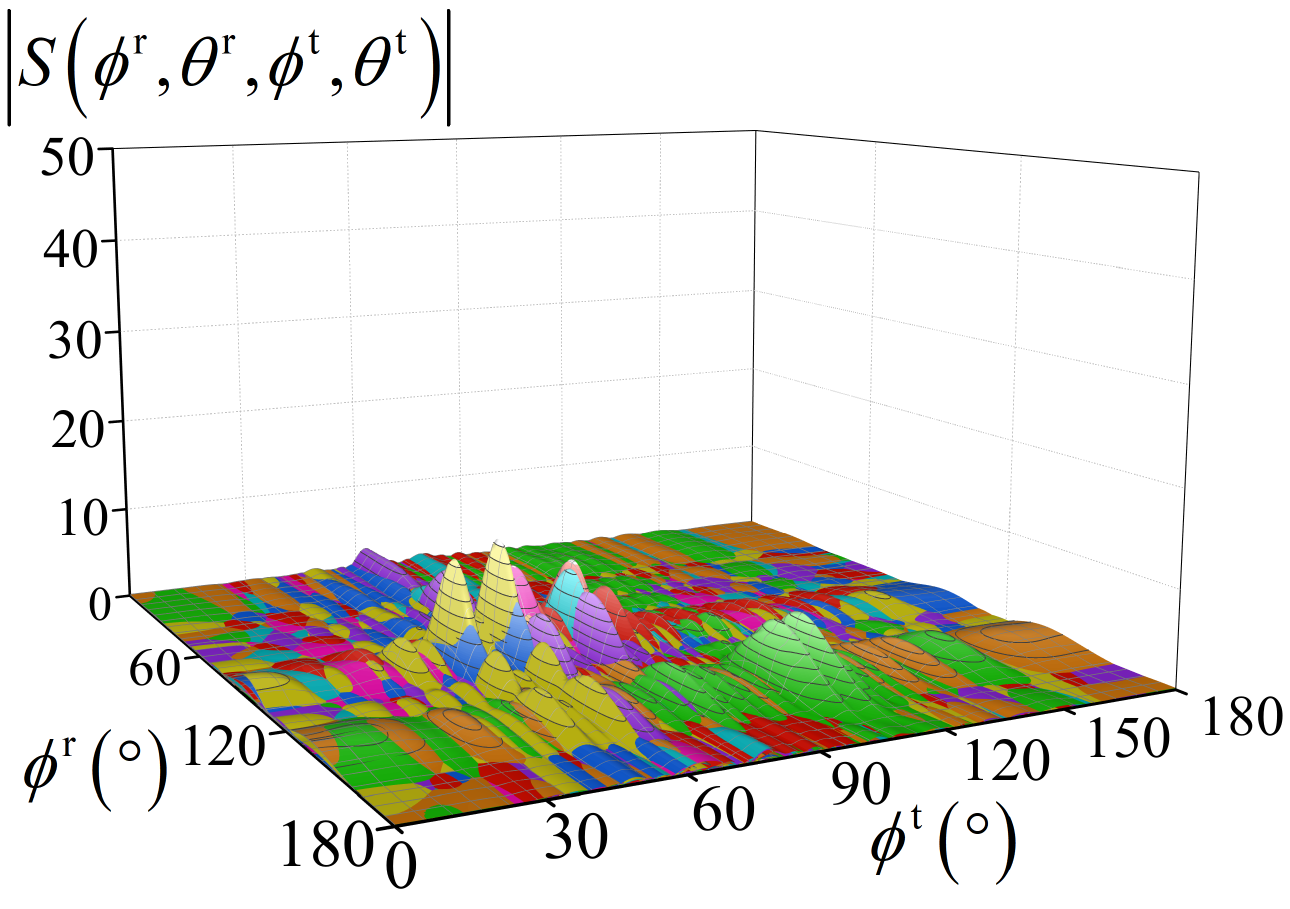}}

  \caption{(a), (b) and (c) are beam patterns of the equivalent channel obtained by different beamforming algorithms, where (e) and (f) are beam patterns of the channel gap compared to the optimal equivalent channel in (a). (d) is a beam pattern of the equivalent channel for which the channel singular vector corresponding to the maximal singular value is served as a precoder. The $z$-axis uses a linear scale.}
  \label{fig3}
\end{figure*}

\begin{algorithm}[t]
  \caption{PE-SMD hybrid beamforming algorithm}\label{algorithm5}
  \begin{algorithmic}[1]
      \Require{${{\mathbf{F}}_{\mathrm{opt}}}$, ${{\mathbf{A}}_{\mathrm{t}}}$, $\mathbf{\tilde{P}\tilde{V}}$}
      \State
      $\mathcal{F}=\left\{ {{l}_{1}},\ldots ,{{l}_{{{{N}_{\mathrm{\!R\!F\!}}^{\mathrm{t}}-{{N}_{\mathrm{\!s}}}}}}} \right\}=\underset{l=1,\ldots ,{{N}_{\mathrm{cl}}}{{N}_{\mathrm{ray}}}}{\mathop{\arg \max }}\,{{\left\| {{\left[ \mathbf{\tilde{P}\tilde{V}} \right]}_{l,1:{{N}_{\mathrm{s}}}}} \right\|}_{2}}$;
      \State 
      ${{\mathbf{P}}_{\mathrm{pjt}}}={{\left( {{\left[ {{\mathbf{A}}_{\mathrm{t}}} \right]}_{:,\mathcal{F}}} \right)}^{H}}{{\mathbf{F}}_{\mathrm{opt}}}$;
      \State
      ${{\mathbf{F}}_{\mathrm{res}}}={{\mathbf{F}}_{\mathrm{opt}}}-{{\left[ {{\mathbf{A}}_{\mathrm{t}}} \right]}_{:,\mathcal{F}}}{{\mathbf{P}}_{\mathrm{pjt}}}$;
      \State
      ${{\mathbf{F}}_{\mathrm{RF}}^{\mathrm{initial}}}=\frac{1}{\sqrt{{{N}_{\mathrm{t}}}}}\exp \left( \mathrm{j}\cdot \arg \left( \left[ {{\mathbf{F}}_{\mathrm{res}}}\ ,\ {{\left[ {{\mathbf{A}}_{\mathrm{t}}} \right]}_{:,\mathcal{F}}}  \right] \right) \right)$;
      \State
      ${{\mathbf{F}}_{\mathrm{BB}}}=\mathbf{F}_{\mathrm{RF}}^{\dagger }{{\mathbf{F}}_{\mathrm{opt}}},\quad {{\mathbf{F}}_{\mathrm{BB}}^{\mathrm{initial}}}=\frac{\sqrt{{{N}_{\mathrm{s}}}}}{{{\left\| {{\mathbf{F}}_{\mathrm{RF}}}{{\mathbf{F}}_{\mathrm{BB}}} \right\|}_{F}}}{{\mathbf{F}}_{\mathrm{BB}}}$;
      \Ensure
      ${{\mathbf{F}}_{\mathrm{RF}}^{\mathrm{initial}}}$, ${{\mathbf{F}}_{\mathrm{BB}}^{\mathrm{initial}}}$
  \end{algorithmic}
\end{algorithm}

\vspace{-5pt}

\subsection{Beam Patterns of Equivalent Channel after Beamforming}
To provide an intuitive explanation of why PE-OMP outperforms the traditional OMP algorithm, this subsection introduces the equivalent channel after beamforming and defines its corresponding beam pattern. Specifically, denote the generic transmitter precoder and receiver combiner as $\mathbf{F}\in {{\mathbb{C}}^{{{N}_{\mathrm{t}}}\times {{N}_{\mathrm{s}}}}}$ and $\mathbf{W}\in {{\mathbb{C}}^{{{N}_{\mathrm{r}}}\times {{N}_{\mathrm{s}}}}}$, respectively. The algorithms discussed in this paper achieve near-optimal solutions, as a result, $\mathbf{W}$ and $\mathbf{F}$ are approximately semi-unitary and satisfy ${{\mathbf{W}}^{H}}\mathbf{W}\approx {{\mathbf{I}}_{{{N}_{\mathrm{s}}}}}$ and ${{\mathbf{F}}^{H}}\mathbf{F}\approx {{\mathbf{I}}_{{{N}_{\mathrm{s}}}}}$. Thus, the beamforming for channel $\mathbf{H}\in {{\mathbb{C}}^{{{N}_{\mathrm{r}}}\times {{N}_{\mathrm{t}}}}}$ can be expressed as
\begin{equation}
  \begin{aligned}
  & {{\mathbf{W}}^{H}}\mathbf{HF}\approx {{\mathbf{W}}^{H}}\left( \mathbf{W}{{\mathbf{W}}^{H}}\mathbf{HF}{{\mathbf{F}}^{H}} \right)\mathbf{F} \\ 
 & \quad \quad \quad \ \,={{\mathbf{W}}^{H}}{{\mathbf{H}}_{\mathrm{eq}}}\mathbf{F}, \\
\end{aligned}
\label{eq44}
\end{equation}
\begin{equation} 
{{\mathbf{H}}_{\mathrm{eq}}}=\mathbf{W}{{\mathbf{W}}^{H}}\mathbf{HF}{{\mathbf{F}}^{H}}\in {{\mathbb{C}}^{{{N}_{\mathrm{r}}}\times {{N}_{\mathrm{t}}}}}, 
\label{eq45}
\end{equation}
where ${{\mathbf{H}}_{\mathrm{eq}}}$ represents the equivalent channel including components of $\mathbf{H}$ that are actually utilized in the communication system after beamforming, while the remaining channel components in $\mathbf{H}-{{\mathbf{H}}_{\mathrm{eq}}}$ are not utilized in actual communication. Therefore, the beam pattern of ${{\mathbf{H}}_{\mathrm{eq}}}$ reflects the effectiveness of the beamforming algorithm in the utilization of the channel and, considering the angles $\left( {{\phi }^{\mathrm{r}}},{{\theta }^{\mathrm{r}}},{{\phi }^{\mathrm{t}}},{{\theta }^{\mathrm{t}}} \right)$, can be defined as
\begin{equation}
S\left( {{\phi }^{\mathrm{r}}},{{\theta }^{\mathrm{r}}},{{\phi }^{\mathrm{t}}},{{\theta }^{\mathrm{t}}} \right)={{\mathbf{a}}_{\mathrm{r}}}{{\left( {{\phi }^{\mathrm{r}}},{{\theta }^{\mathrm{r}}} \right)}^{H}}{{\mathbf{H}}_{\mathrm{eq}}}{{\mathbf{a}}_{\mathrm{t}}}\left( {{\phi }^{\mathrm{t}}},{{\theta }^{\mathrm{t}}} \right).
\label{eq46}
\end{equation}

When the fully-digital optimal algorithm is employed, its equivalent channel can maximize the spectral efficiency. The optimal equivalent channel can be expressed as
\begin{equation}
  \mathbf{H}_{\mathrm{eq}}^{\mathrm{opt}}={{\mathbf{W}}_{\mathrm{opt}}}\mathbf{W}_{\mathrm{opt}}^{H}\mathbf{H}{{\mathbf{F}}_{\mathrm{opt}}}\mathbf{F}_{\mathrm{opt}}^{H}.
\label{eq47}
\end{equation}
The ``channel gap" between $\mathbf{H}_{\mathrm{eq}}^{\mathrm{opt}}$ and the equivalent channels of other algorithms can be expressed as
\begin{equation}
  {{\mathbf{H}}_{\mathrm{gap}}}=\mathbf{H}_{\mathrm{eq}}^{\mathrm{opt}}-{{\mathbf{H}}_{\mathrm{eq}}}.
  \label{eq48}
\end{equation}
The smaller ${{\left\| {{\mathbf{H}}_{\mathrm{gap}}} \right\|}_{F}}$, the better performance of the beamforming algorithm. Therefore, the performance differences among various algorithms can be intuitively illustrated through the beam patterns of ${{\mathbf{H}}_{\mathrm{eq}}}$ and ${{\mathbf{H}}_{\mathrm{gap}}}$, which will be shown in Sec.~VII. In the following section, the algorithms are analyzed from the complexity point of view.

\begin{table}[t]
  \renewcommand{\arraystretch}{1.5} 
  \centering
  \caption{{\textsc{\\Computational complexity of the proposed algorithms}}}\label{tab1}
  \begin{tabular}{|l|l|}
  \hline
  \textbf{Algorithm} & \textbf{Complexity} \\
  \hline

  AREE &
  $\mathcal{O}\left( \left( 4{{N}_{\mathrm{t}}}N_{\mathrm{s}}^{2}\!{{N}_{\mathrm{\!iter\!1}}}\!+\!2{{N}_{\mathrm{t}}}N_{\mathrm{\!R\!F}}^{\mathrm{t}}\!\left( N_{\mathrm{\!R\!F}}^{\mathrm{t}}\!-\!{{N}_{\mathrm{s}}}\! \right)\!{{N}_{\mathrm{iter2}}} \right)\!{{N}_{\mathrm{iter3}}} \right)$ \\
  \hline

  GC-SVD &
  $\mathcal{O}\left( \left( {{N}_{\mathrm{t}}}+{{N}_{\mathrm{r}}} \right){{\left( {{N}_{\mathrm{cl}}}{{N}_{\mathrm{ray}}} \right)}^{2}} \right)$ \\
  \hline

  PE-OMP &
  $\mathcal{O}\left( {{N}_{\mathrm{t}}}{{N}_{\mathrm{s}}}{{N}_{\mathrm{cl}}}{{N}_{\mathrm{ray}}}\!\left( N_{\mathrm{\!RF}}^{\mathrm{t}}\!-\!{{N}_{\mathrm{s}}} \!\right)\!+\!{{N}_{\mathrm{t}}}N_{\mathrm{\!RF}}^{\mathrm{t}}\!\left( 2N_{\mathrm{\!RF}}^{\mathrm{t}}\!+\!{{N}_{\mathrm{s}}} \right) \right)$ \\
  \hline

  PE-SMD &
  $\mathcal{O}\left( {{N}_{\mathrm{t}}}N_{\mathrm{RF}}^{\mathrm{t}}\left( 2N_{\mathrm{RF}}^{\mathrm{t}}+{{N}_{\mathrm{s}}} \right) \right)$ \\
  \hline
  \end{tabular}
\end{table}

\vspace{-5pt}

\section{Complexity Analysis}

The computational complexity of the proposed algorithms is summarized in Table \ref{tab1}, where ${{N}_{\mathrm{iter1}}}$, ${{N}_{\mathrm{iter2}}}$ and ${{N}_{\mathrm{iter3}}}$ represent the number of $iterations1$, $iterations2$ and $iterations3$ defined in \textbf{Algorithm 1}, respectively. Among iterative hybrid beamforming algorithms, PE-AltMin \cite{yu2016alternating} exhibits a complexity of $\mathcal{O}\left( {{N}_{\mathrm{t}}}N_{\mathrm{RF}}^{\mathrm{t}}\left( 2N_{\mathrm{RF}}^{\mathrm{t}}+3{{N}_{\mathrm{s}}} \right){{N}_{\mathrm{iter1}}} \right)$, but it suffers from a significant performance loss when $N_{\mathrm{RF}}^{\mathrm{t}}>{{N}_{\mathrm{s}}}$. When ${{N}_{\mathrm{iter3}}=1}$, the proposed AREE algorithm decomposes the original high-dimensional problem into two low-dimensional subproblems, providing a complexity comparable to that of PE-AltMin. When ${{N}_{\mathrm{iter3}}> 1}$, AREE outperforms existing algorithms by alternately optimizing two subproblems iteratively, while each iteration converges quickly, thus keeping ${{N}_{\mathrm{iter1}}}$ and ${{N}_{\mathrm{iter2}}}$ small (typically no more than 2, as shown in Sec.~VII). When PE-SMD serves as an initial value, AREE converges rapidly, therefore keeping computational complexity comparable to that of PE-AltMin.

Traditional SVD algorithms for mmWave XL-MIMO channels exhibit a relatively high complexity of $\mathcal{O}\left( {{N}_{\mathrm{t}}}N_{\mathrm{r}}^{2} \right)$. In contrast, the GC-SVD algorithm decomposes the high-dimensional SVD into three low-dimensional SVDs by exploiting the mmWave channel sparsity, which significantly reduces complexity particularly in line-of-sight scenarios such as vehicular networks (which satisfy ${{N}_{\mathrm{r}}}>{{N}_{\mathrm{cl}}}{{N}_{\mathrm{ray}}}$).

Among non-iterative hybrid beamforming algorithms, the existing OMP-based algorithm \cite{el2014spatially} provides a slightly high complexity of $\mathcal{O}\left( {{N}_{\mathrm{t}}}{{N}_{\mathrm{s}}}{{N}_{\mathrm{cl}}}{{N}_{\mathrm{ray}}}N_{\mathrm{\!RF}}^{\mathrm{t}}\!+\!{{N}_{\mathrm{t}}}N_{\mathrm{\!RF}}^{\mathrm{t}}\left( 2N_{\mathrm{\!RF}}^{\mathrm{t}}\!+\!{{N}_{\mathrm{s}}} \right) \right)$. In contrast, the PE-OMP algorithm first selects $N_{\mathrm{RF}}^{\mathrm{t}}\!-\!{{N}_{\mathrm{s}}}$ strongest beam components, and then retains the remaining beam components by a simple phase extraction operation using the remaining ${{N}_{\mathrm{s}}}$ RF chains. This not only reduces complexity but also improves performance by taking into account all beam components. The PE-SMD algorithm simplifies PE-OMP by avoiding projection and pseudo-inverse operations, achieving lower complexity while incurring only slight performance loss.

\section{Simulation Results}
In this section, we present extensive simulation results to validate the performance of the proposed algorithms. For comparison, we include both existing non-iterative algorithms (OMP \cite{el2014spatially} and SS-SVD \cite{tsai2018sub}) and iterative algorithms (PE-AltMin \cite{yu2016alternating}, TensDecomp \cite{zilli2021constrained}, and MO-AltMin \cite{yu2016alternating}), along with the optimal fully-digital solution. Both the transmitter and receiver employ USPAs with ${{N}_{\mathrm{t}}}=256$ and ${{N}_{\mathrm{r}}}=64$ antennas, respectively. The number of data streams being transmitted is ${{N}_{\mathrm{s}}}=6$, while the number of RF chains satisfies ${{N}_{\mathrm{s}}}\le N_{\mathrm{RF}}^{\mathrm{t}}=N_{\mathrm{RF}}^{\mathrm{r}}<2{{N}_{\mathrm{s}}}$. For the mmWave channels, we assume ${{N}_{\mathrm{cl}}}=5$, ${{N}_{\mathrm{ray}}}=10$ and ${{\sigma }_{\mathrm{\phi} }}={{\sigma }_{\mathrm{\theta} }}={{10}^{\circ }}$ as described in \cite{akdeniz2014millimeter,yu2016alternating}, and other parameters, including ${{\alpha }_{i,l}}$, $\phi _{i,l}^{\mathrm{t}}$, $\phi _{i,l}^{\mathrm{r}}$, $\theta _{i,l}^{\mathrm{t}}$ and $\theta _{i,l}^{\mathrm{r}}$ as specified in Sec.~II. All simulation results are averaged over 1000 channel realizations, except for the single channel realization used in Fig.~\ref{fig3}. The numbers of $iterations1$, $iterations2$ and $iterations3$ defined in \textbf{Algorithm 1} are represented as ${{N}_{\mathrm{iter1}}}$, ${{N}_{\mathrm{iter2}}}$ and ${{N}_{\mathrm{iter3}}}$, respectively. The signal-to-noise ratio (SNR) used in the simulations is defined as
\begin{equation}
  \text{SNR}=\frac{\operatorname{Tr}\left( \mathbb{E}\left[ \mathbf{r}{{\mathbf{r}}^{H}} \right] \right)}{\operatorname{Tr}\left( \mathbb{E}\left[ \mathbf{\tilde{n}}{{{\mathbf{\tilde{n}}}}^{H}} \right] \right)}=\frac{{{P}_{\mathrm{t}}}}{{{N}_{\mathrm{s}}}\sigma _{\mathrm{n}}^{2}}\frac{\left\| \mathbf{W}_{\mathrm{BB}}^{H}\mathbf{W}_{\mathrm{RF}}^{H}\mathbf{H}{{\mathbf{F}}_{\mathrm{RF}}}{{\mathbf{F}}_{\mathrm{BB}}} \right\|_{F}^{2}}{\left\| \mathbf{W}_{\mathrm{BB}}^{H}\mathbf{W}_{\mathrm{RF}}^{H} \right\|_{F}^{2}},
\label{eq49}
\end{equation}
where $\mathbf{r}=\mathbf{W}_{\mathrm{BB}}^{H}\mathbf{W}_{\mathrm{RF}}^{H}\mathbf{H}{{\mathbf{F}}_{\mathrm{RF}}}{{\mathbf{F}}_{\mathrm{BB}}}\mathbf{s}$ and $\mathbf{\tilde{n}}=\mathbf{W}_{\mathrm{BB}}^{H}\mathbf{W}_{\mathrm{RF}}^{H}\mathbf{n}$ are reformulations of the variables in (\ref{eq1}).

\subsection{Comparison of Beam Patterns for Non-Iterative Algorithms}

Fig. \ref{fig3} shows the beam patterns $S\left( {{\phi }^{\mathrm{r}}},{{\theta }^{\mathrm{r}}},{{\phi }^{\mathrm{t}}},{{\theta }^{\mathrm{t}}} \right)$ defined in (\ref{eq46}) of the equivalent channels $\mathbf{H}_{\mathrm{eq}}$ (see (\ref{eq45})) and channel gaps $\mathbf{H}_{\mathrm{gap}}$ (see (\ref{eq48})) obtained by different beamforming algorithms. Since $S\left( {{\phi }^{\mathrm{r}}},{{\theta }^{\mathrm{r}}},{{\phi }^{\mathrm{t}}},{{\theta }^{\mathrm{t}}} \right)$ depends on four variables, we represent the beam pattern within a three-dimensional space by assigning $\left| S\left( {{\phi }^{\mathrm{r}}},{{\theta }^{\mathrm{r}}},{{\phi }^{\mathrm{t}}},{{\theta }^{\mathrm{t}}} \right) \right|$, azimuth angles ${{\phi }^{\mathrm{r}}}$ and ${{\phi }^{\mathrm{t}}}$ to the $z$-axis, $x$-axis, and $y$-axis, respectively. The beam patterns at different elevation angles $\left( {{\theta }^{\mathrm{r}}},{{\theta }^{\mathrm{t}}} \right)$ are displayed in the 3D coordinate system with different colors, as illustrated in Fig. \ref{fig3}, where the values of $\left( {{\theta }^{\mathrm{r}}},{{\theta }^{\mathrm{t}}} \right)$ are sampled from ${{\theta }^{r}},{\theta }^{t}\in \left\{ {\pi }/{6}\;+\left( {\pi }/{12}\; \right)\cdot i \right\}_{i=0}^{8}$.

Compared to the fully-digital optimal beam pattern in Fig.~\ref{fig3}\subref{fig3a}, the OMP algorithm, as depicted in Fig.~\ref{fig3}\subref{fig3b}, selects only $N_{RF}^{\mathrm{t}}$ strongest beam components from Fig.~\ref{fig3}\subref{fig3a}, while discarding all the remaining beam components. The discarded components form the ``channel gap" illustrated in Fig.~\ref{fig3}\subref{fig3e}. It is evident that the channel gap of OMP has considerable energy, which may even exceed the energy retained in Fig.~\ref{fig3}\subref{fig3b} when $N_{\mathrm{RF}}^{\mathrm{t}}$ is relatively small, leading to significant performance loss. In contrast, the proposed PE-OMP algorithm in Fig.~\ref{fig3}\subref{fig3c} first selects $N_{\mathrm{RF}}^{\mathrm{t}}-{{N}_{\mathrm{s}}}$ strongest beam components from Fig.~\ref{fig3}\subref{fig3a}, and then captures all remaining beam components using the remaining ${{N}_{\mathrm{s}}}$ RF chains. Taking into account all the beam components to reconstruct precoders, Fig.~\ref{fig3}\subref{fig3c} more closely approximates Fig.~\ref{fig3}\subref{fig3a} with a reduced channel gap in Fig.~\ref{fig3}\subref{fig3f}, resulting in enhanced performance.

Additionally, Fig.~\ref{fig3}\subref{fig3d} shows the beam pattern of the effective channel, {\small ${{\left[ \mathbf{U} \right]}_{:,1}}{{\left( {{\left[ \mathbf{U} \right]}_{:,1}} \right)}^{H}}\mathbf{H}{{\left[ \mathbf{V} \right]}_{:,1}}{{\left( {{\left[ \mathbf{V} \right]}_{:,1}} \right)}^{H}}$}, where the channel singular vector corresponding to the maximum singular value is used as a precoder. This beam pattern is observed to be a linear combination of multiple beams (steering vectors) from different directions, further validating the relationships described in (\ref{eq39}) and (\ref{eq40}).

\begin{figure}[t]
  \centering
  \includegraphics[width=0.9\columnwidth]{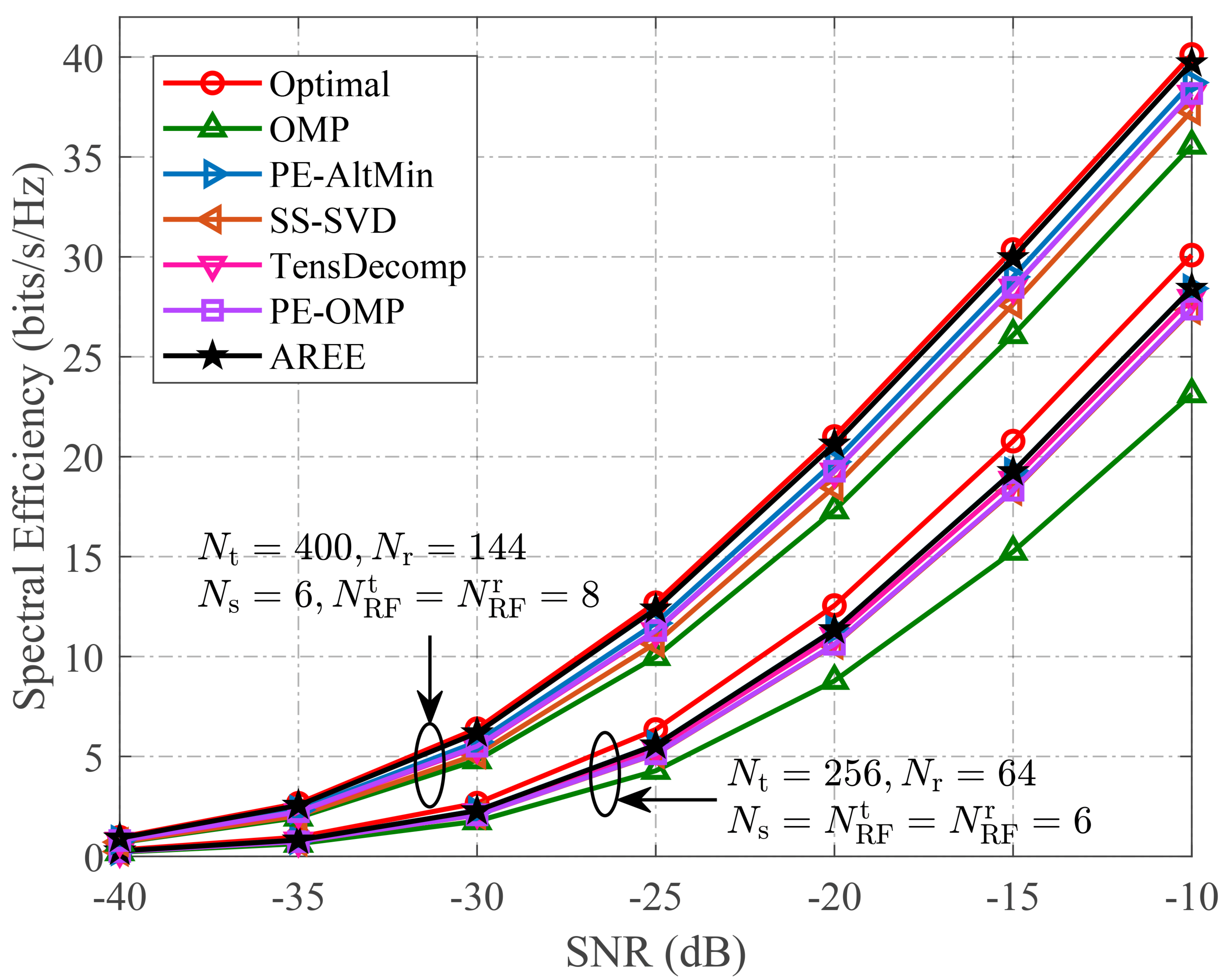}
  \caption{Relationships between spectral efficiency and SNR for different beamforming algorithms, where initial values of the AREE algorithm are randomly assigned.}
  \label{fig4}
\end{figure}
\begin{figure}[t]
  \centering
  \includegraphics[width=0.9\columnwidth]{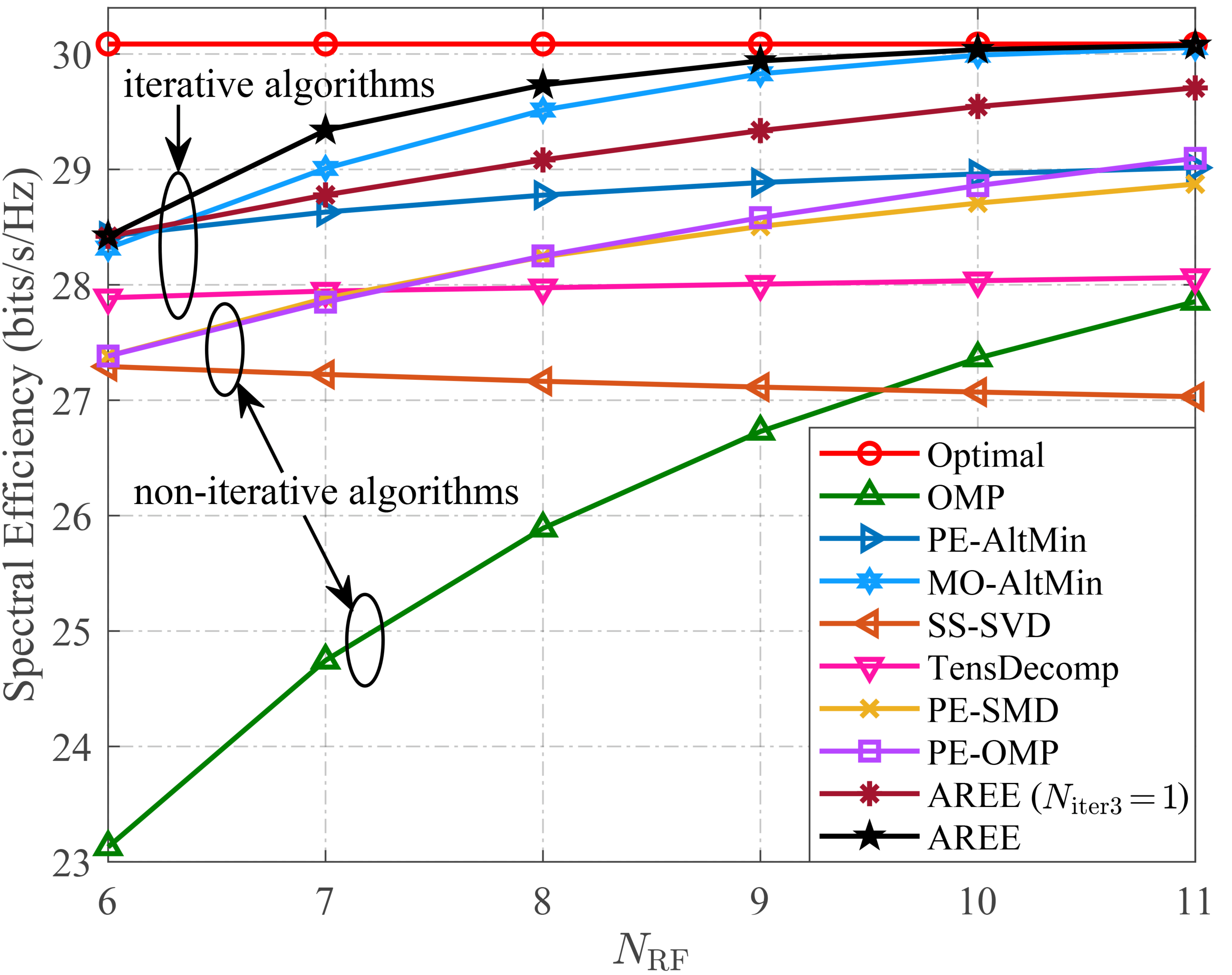}
  \caption{Relationships between spectral efficiency and the number of RF chains for different beamforming algorithms, where initial values of the AREE are randomly assigned. ${{N}_{\mathrm{s}}}=6$, $N_{\mathrm{RF}}=N_{\mathrm{RF}}^{\mathrm{r}}=N_{\mathrm{RF}}^{\mathrm{t}}$ and $\text{SNR}=-10\ \text{dB}$.}
  \label{fig5}
\end{figure}

\begin{figure}[t]
  \raggedright 
  \includegraphics[width=0.9175\columnwidth]
  {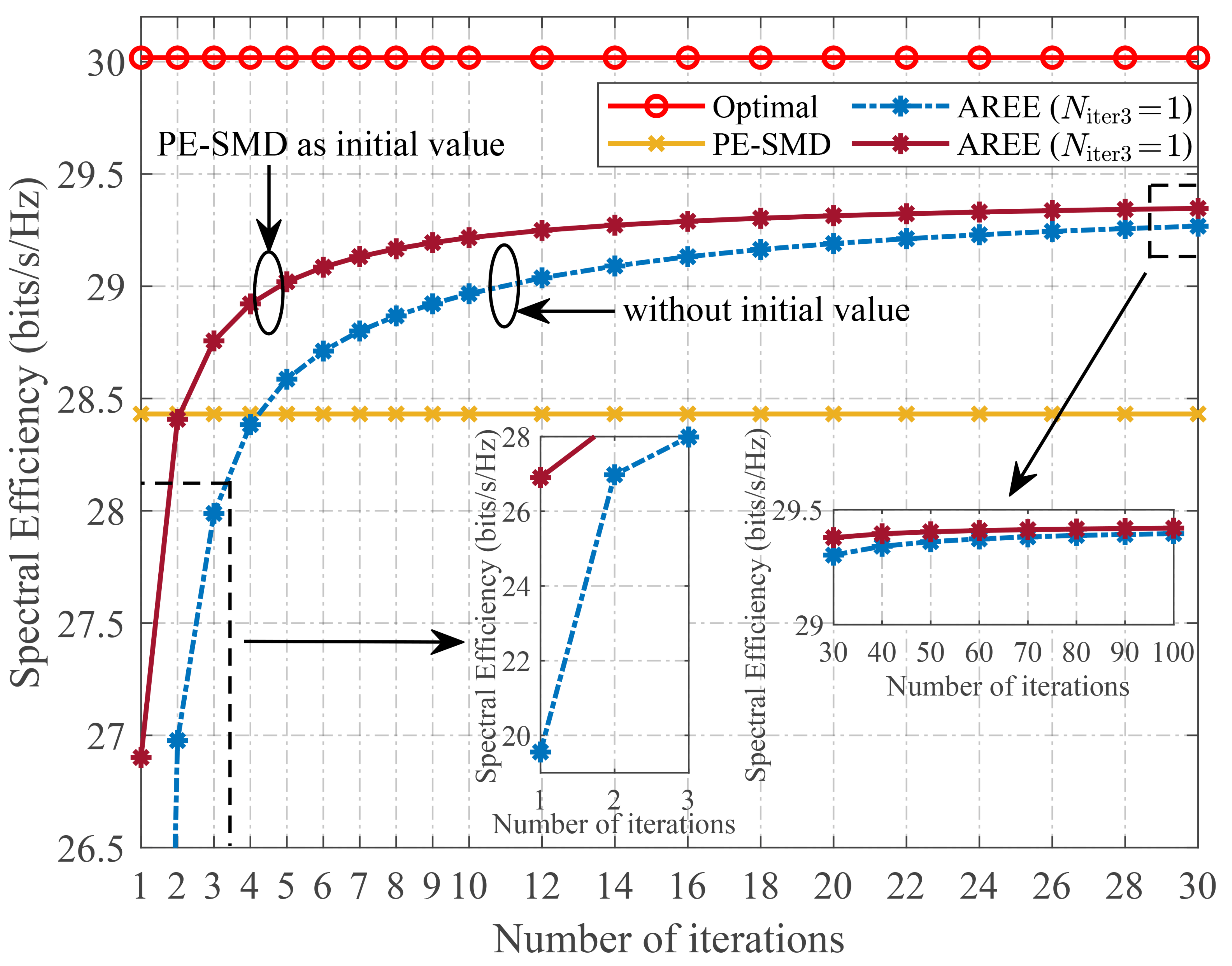}
  \caption{Convergence speeds of $iterations1$ and $iterations2$ in AREE algorithm when ${{N}_{\mathrm{iter3}}}=1$. The $x$-axis variable represents ${{N}_{\mathrm{iter1}}}={{N}_{\mathrm{iter2}}}$. ${{N}_{\mathrm{s}}}=6$, $N_{\mathrm{RF}}^{\mathrm{t}}=N_{\mathrm{RF}}^{\mathrm{r}}=9$ and $\text{SNR}=-10\ \text{dB}$.}
  \label{fig6}
\end{figure}

\begin{figure}[t]
  \hspace{0.04cm}
  \includegraphics[width=0.98\columnwidth]{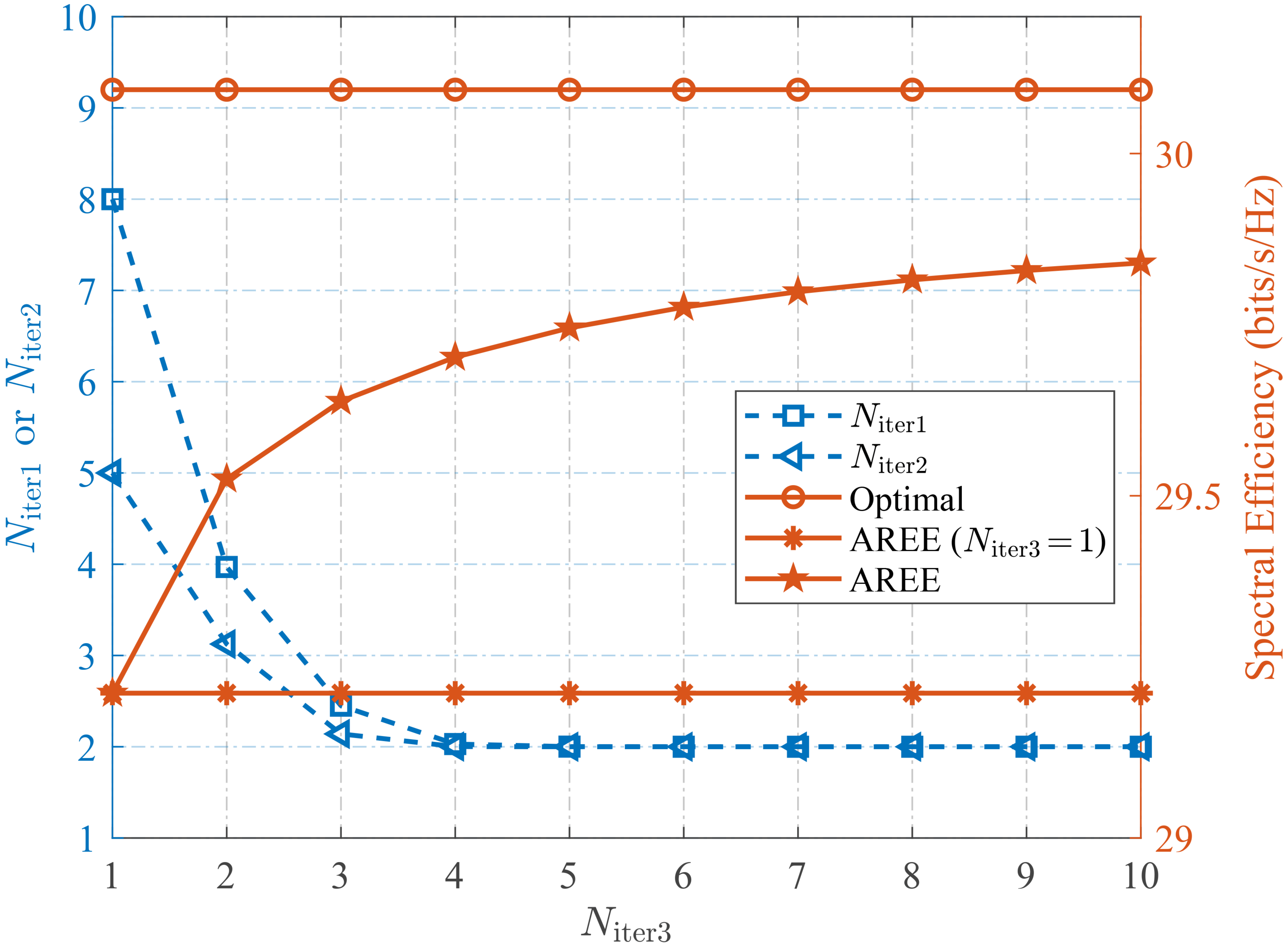}
  \caption{${{N}_{\mathrm{iter1}}}$ and ${{N}_{\mathrm{iter2}}}$ required for convergence within each $iterations3$, where PE-SMD is served as an initial value for the AREE. ${{N}_{\mathrm{s}}}=6$, $N_{\mathrm{RF}}^{\mathrm{t}}=N_{\mathrm{RF}}^{\mathrm{r}}=9$ and $\text{SNR}=-10\ \text{dB}$.}
  \label{fig7}
\end{figure}

\subsection{Comparison of Spectral Efficiency}
Fig. \ref{fig4} illustrates the spectral efficiency versus SNR for two different antenna and RF chain configurations. The proposed AREE algorithm demonstrates superior performance over all baseline methods, particularly when $N_{\mathrm{RF}}>{{N}_{\mathrm{s}}}$. This advantage stems from its effective precoder decoupling mechanism, which simultaneously improves both computational and spectral efficiency. Among non-iterative approaches, the proposed PE-OMP surpasses both OMP and SS-SVD by comprehensively incorporating all effective beam components in precoder reconstruction, while maintaining lower computational complexity than OMP through fewer orthogonal projections.

Fig. \ref{fig5} shows the spectral efficiency as a function of the number of RF chains. For non-iterative algorithms, while the proposed PE-SMD shows marginally inferior performance to PE-OMP, both proposed methods significantly outperform existing OMP and SS-SVD. Notably, SS-SVD performance degrades with increasing $N_{\mathrm{RF}}$ due to its inclusion of ineffective beam components. Among iterative algorithms, the proposed AREE algorithm closely approximates or even achieves the optimal solution as $N_{\mathrm{RF}}$ increases, achieving the best performance among all compared methods with maintained low complexity. This superior performance results from its effective decoupling of analog and digital precoders. In contrast, existing algorithms such as PE-AltMin and TensDecomp fail  to effectively optimize RF chains as $N_{\mathrm{RF}}$ increases, resulting in almost negligible performance improvements. Meanwhile, MO-AltMin's direct optimization of the coupled precoders incurs substantially higher computational overhead.

\subsection{The Convergence Speed of Iterative Algorithms}
Fig. \ref{fig6} shows the convergence behaviors of $iterations1$ and $iterations2$ in AREE algorithm under different initial values when ${{N}_{\mathrm{iter3}}}=1$. Using PE-SMD as an initial value, convergence is much faster than for the AREE algorithm with a randomly assigned initial value (i.e., without specific initialization). This indicates that, for ${{N}_{\mathrm{iter3}}}=1$, the computational complexity of AREE is further reduced and becomes comparable to that of PE-AltMin, while achieving improved spectral efficiency.

\begin{figure}[t]
  \raggedright 
  \includegraphics[width=0.9\columnwidth]{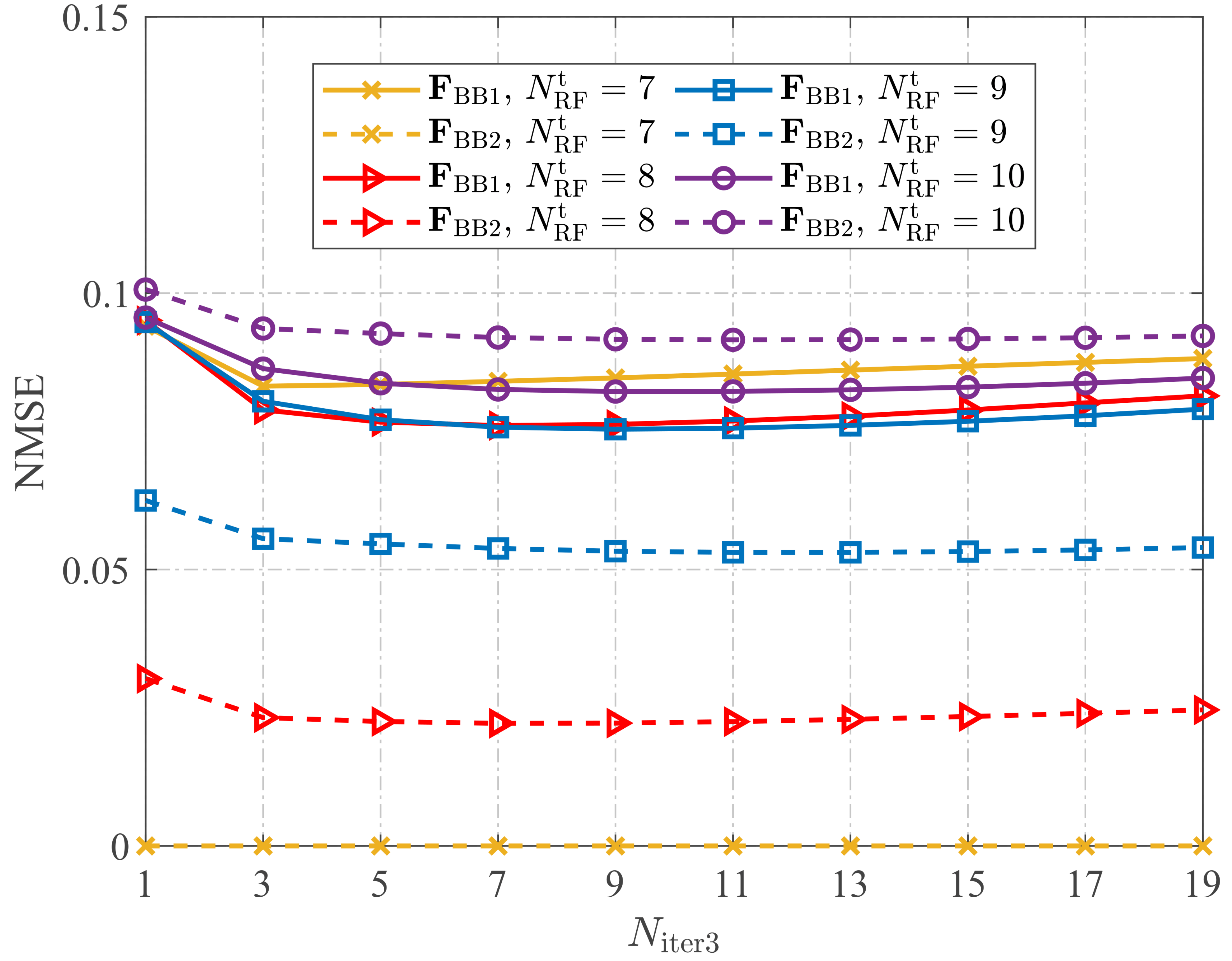}
  \caption{The normalized mean squared errors $\text{NMSE}\!\left( {{\mathbf{F}}_{\mathrm{\!BB1}}} \!\right)$ and $\text{NMSE}\!\left( {{\mathbf{F}}_{\mathrm{\!BB2}}} \!\right)$, as defined in (\ref{eq30}), are evaluated at the convergence of each $iterations3$. Initial values of the AREE algorithm are randomly assigned. ${{N}_{\mathrm{s}}}=6$, $N_{\mathrm{RF}}^{\mathrm{r}}=N_{\mathrm{RF}}^{\mathrm{t}}$ and $\text{SNR}=-10\ \text{dB}$.}
  \label{fig8}
\end{figure}

The convergence behavior at each step of $iterations3$ in AREE algorithm is further analyzed in Fig. \ref{fig7}, where PE-SMD serves as the initial value. The orange curves represent the spectral efficiency achieved after each $iterations3$ converges, while the blue curves depict ${{N}_{\mathrm{iter1}}}$ and ${{N}_{\mathrm{iter2}}}$ required for convergence within each $iterations3$. When ${{N}_{\mathrm{iter3}}}\!=\!1$, AREE converges with ${{N}_{\mathrm{iter1}}}\!=\!8$ and ${{N}_{\mathrm{iter2}}}\!=\!5$, which is consistent with the results in Fig. \ref{fig6}. For ${{N}_{\mathrm{iter3}}}\!\ge\! 3$, ${{N}_{\mathrm{iter1}}}\!=\!{{N}_{\mathrm{iter2}}}\!=\!2$ is sufficient for rapid convergence, as the performance is already close to the optimal solution, requiring only minor error corrections in each iteration to achieve fast re-convergence. Since ${{N}_{\mathrm{iter3}}}\le 10$ is typically near convergence, AREE maintains complexity comparable to PE-AltMin, while outperforming existing high-complexity algorithms such as MO-AltMin in spectral efficiency. This indicates that the AREE algorithm achieves enhanced performance with reduced complexity.

\subsection{Analysis of the Optimal Conditions and Matrix Partition}
Fig. \ref{fig8} illustrates the similarity between ${{\mathbf{F}}_{\mathrm{BB1}}}$, ${{\mathbf{F}}_{\mathrm{BB2}}}$ and the semi-unitary matrix during the evolution of the AREE algorithm, evaluated using the NMSE metric defined in (\ref{eq30}). Under various configurations of ${N}_{\mathrm{RF}}^{\mathrm{t}}$, both $\text{NMSE}\left( {{\mathbf{F}}_{\mathrm{BB1}}} \right)$ and $\text{NMSE}\left( {{\mathbf{F}}_{\mathrm{BB2}}} \right)$ remain below 0.1 and exhibit minimal variation across $iterations3$. This indicates that throughout the AREE algorithm, ${{\mathbf{F}}_{\mathrm{BB1}}}$ and ${{\mathbf{F}}_{\mathrm{BB2}}}$ maintain a high degree of similarity to the semi-unitary matrix. Consequently, as derived from (\ref{eq29}), (\ref{eq19}) approximates the optimal solution of the first subproblem (\ref{eq14}). Similarly, as ${N}_{\mathrm{RF}}^{\mathrm{t}
}$ increases, ${{\mathbf{F}}_{\mathrm{BB2}}}$ approaches a square matrix, allowing (\ref{eq23}) to approximate the optimal solution of the second subproblem (\ref{eq21}). When the number of RF chains approaches $2{{N}_{\mathrm{s}}}$, both subproblems nearly satisfy their respective optimality conditions. Therefore, as demonstrated in Fig. \ref{fig5}, the AREE algorithm achieves the optimal solution.

\begin{figure}[t]
  \raggedright 
  \includegraphics[width=0.91\columnwidth]{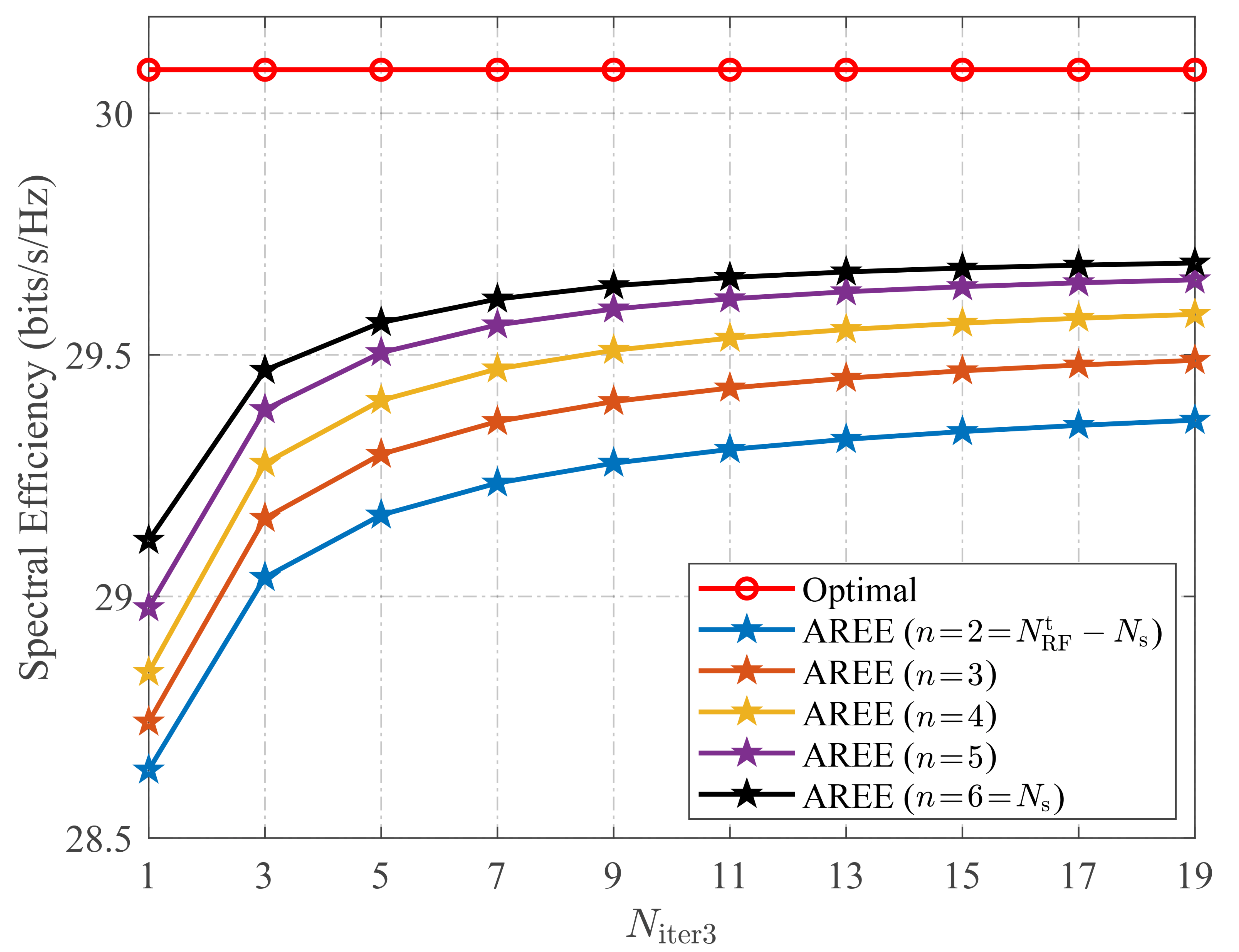}
  \caption{The spectral efficiency of different matrix partition schemes, as defined in (\ref{eq31}) and (\ref{eq32}), is evaluated at the convergence of each $iterations3$. Intial values of the AREE algorithm are randomly assigned. ${{N}_{\mathrm{s}}}=6$, $N_{\mathrm{RF}}^{\mathrm{t}}=N_{\mathrm{RF}}^{\mathrm{r}}=8$ and $\text{SNR}=-10\ \text{dB}$.}
  \label{fig9}
\end{figure}

Fig. \ref{fig9} shows the spectral efficiency of the AREE algorithm under different matrix partition schemes for ${{\mathbf{F}}_{\mathrm{RF}}}$ and ${{\mathbf{F}}_{\mathrm{BB}}}$, as defined in (\ref{eq31}) and (\ref{eq32}). When $n={{N}_{\mathrm{s}}}$, ${{\mathbf{F}}_{\mathrm{BB1}}}$ becomes a square matrix, enabling the first subproblem (\ref{eq14}) to achieve better optimization results in the initial stage. This provides a higher-quality initial value for the subsequent alternate iterations, which ultimately results in improved overall performance. Conversely, as $n$ decreases, the spectral efficiency deteriorates, indicating that $n={{N}_{\mathrm{s}}}$ is the optimal matrix partition scheme for spectral efficiency, as it ensures a more effective initial value for iteration. Although providing a theoretical proof of this optimality lies beyond the scope of this paper, it remains an intriguing topic for future research.

\section{Conclusion}
This paper investigated mmWave hybrid beamforming with decoupled precoders in XL-MIMO systems. The proposed AREE algorithm decomposes the hybrid beamforming problem into two low-dimensional subproblems, which decouple precoders with negligible errors. These subproblems alternately eliminate each other's residual errors, driving the overall solution toward optimal hybrid beamforming performance. Additionally, the proposed GC-SVD algorithm facilitates the derivation of subproblems by transforming the high-dimensional sparse channel into a low-dimensional representation, while the PE-SMD method provides effective initial values to accelerate convergence of the AREE. Simulation results demonstrate that the AREE algorithm effectively decouples precoders, achieves fast convergence, maintains low complexity, and surpasses existing algorithms in spectral efficiency. In future works, partially connected hybrid array architectures deserve further investigation.

{\appendix
We first prove that if ${{\left\| {\mathbf{E}_{\mathrm{1}}}-\mathbf{F}_{\mathrm{RF2}}^{\mathrm{opt}}\mathbf{F}_{\mathrm{BB2}}^{\mathrm{opt}} \right\|}_{F}}\le \delta $ holds, then ${{\left\| {\mathbf{E}_{\mathrm{1}}}-\mathbf{F}_{\mathrm{RF2}}^{\mathrm{opt}}\mathbf{\hat{F}}_{\mathrm{BB2}}^{\mathrm{opt}} \right\|}_{F}}<2\delta $ also holds, i.e., if $\mathbf{F}_{\mathrm{BB2}}^{\mathrm{opt}}$ approximates the suboptimal solution to the subproblem (\ref{eq21}), $\mathbf{\hat{F}}_{\mathrm{BB2}}^{\mathrm{opt}}$ will also approach it with a similar approximation error.  $\mathbf{\hat{F}}_{\mathrm{BB2}}^{\mathrm{opt}}$ can be further expressed as
\begin{small}
\begin{equation}
\begin{aligned}
  & \mathbf{\hat{F}}_{\mathrm{BB2}}^{\mathrm{opt}}= \frac{\sqrt{{{N}_{\mathrm{s}}}}{{\left\| \mathbf{F}_{\mathrm{RF2}}^{\mathrm{opt}}\mathbf{F}_{\mathrm{BB2}}^{\mathrm{opt}} \right\|}_{F}}}{{{\left\| \mathbf{F}_{\mathrm{RF}}^{\mathrm{opt}}\mathbf{F}_{\mathrm{BB}}^{\mathrm{opt}} \right\|}_{F}}}\frac{1}{{{\left\| \mathbf{F}_{\mathrm{RF2}}^{\mathrm{opt}}\mathbf{F}_{\mathrm{BB2}}^{\mathrm{opt}} \right\|}_{F}}}\mathbf{F}_{\mathrm{BB2}}^{\mathrm{opt}} \\ 
 & \quad \quad  \ =\frac{\sqrt{{{P}_{1}}}}{{{\left\| \mathbf{F}_{\mathrm{RF2}}^{\mathrm{opt}}\mathbf{F}_{\mathrm{BB2}}^{\mathrm{opt}} \right\|}_{F}}}\mathbf{F}_{\mathrm{BB2}}^{\mathrm{opt}} \ ,\\ 
\end{aligned}
\label{eq50}
\end{equation}
\end{small}where {\small $\sqrt{{{P}_{1}}}{=\sqrt{{{N}_{\mathrm{s}}}}{{\left\| \mathbf{F}_{\mathrm{RF2}}^{\mathrm{opt}}\mathbf{F}_{\mathrm{BB2}}^{\mathrm{opt}} \right\|}_{F}}}/{{{\left\| \mathbf{F}_{\mathrm{RF}}^{\mathrm{opt}}\mathbf{F}_{\mathrm{BB}}^{\mathrm{opt}} \right\|}_{F}}}\;$}. Denoting {\small $\sqrt{{{P}_{2}}}={{\left\| {\mathbf{E}_{\mathrm{1}}} \right\|}_{F}}$} and {\small ${1}/{\gamma }\;={\sqrt{{{N}_{\mathrm{s}}}}}/{{{\left\| \mathbf{F}_{\mathrm{RF}}^{\mathrm{opt}}\mathbf{F}_{\mathrm{BB}}^{\mathrm{opt}} \right\|}_{F}}}\;$}, it follows that {\small ${{\left\| \mathbf{F}_{\mathrm{RF2}}^{\mathrm{opt}}\mathbf{F}_{\mathrm{BB2}}^{\mathrm{opt}} \right\|}_{F}}=\gamma \sqrt{{{P}_{1}}}=\gamma \sqrt{{{{P}_{1}}}/{{{P}_{2}}}\;}{{\left\| {\mathbf{E}_{\mathrm{1}}} \right\|}_{F}}$}, and then we have
\begin{small}
\begin{equation}
  {{\left\| {\mathbf{E}_{\mathrm{1}}}-\mathbf{\!F}_{\mathrm{\!RF2}}^{\mathrm{opt}}\mathbf{F}_{\mathrm{\!BB2}}^{\mathrm{opt}} \right\|}_{\!F}}\!\ge \!\left| {{\left\| {\mathbf{E}_{\mathrm{1}}} \right\|}_{\!F}}\!-\!{{\left\| \mathbf{F}_{\mathrm{\!RF2}}^{\mathrm{opt}}\mathbf{F}_{\mathrm{\!BB2}}^{\mathrm{opt}} \right\|}_{\!F}} \right|\!=\!\left| 1\!-\!\gamma \sqrt{\!\frac{{{P}_{1}}}{{{P}_{2}}}} \right|\!{{\left\| {\mathbf{E}_{\mathrm{1}}} \right\|}_{\!F}},
  \label{eq51}
\end{equation}
\end{small}which means {\small ${{\left\| {\mathbf{E}_{\mathrm{1}}} \right\|}_{F}}\le \left( {1}/{\left| 1-\gamma \sqrt{{{{P}_{1}}}/{{{P}_{2}}}\;} \right|}\; \right)\delta $}, and it follows
\begin{small}
\begin{equation}
  \begin{aligned}
  & {{\left\| {\mathbf{E}_{\mathrm{1}}}\!-\!\mathbf{F}_{\mathrm{\!RF2}}^{\mathrm{opt}}\mathbf{\hat{F}}_{\mathrm{\!BB2}}^{\mathrm{opt}} \right\|}_{\!F}}\!=\!{{\left\| {\mathbf{E}_{\mathrm{1}}}\!-\!\mathbf{F}_{\mathrm{\!RF2}}^{\mathrm{opt}}\mathbf{F}_{\mathrm{\!BB2}}^{\mathrm{opt}}+\!\left(\! 1\!-\!\frac{1}{\gamma } \right)\mathbf{F}_{\mathrm{\!RF2}}^{\mathrm{opt}}\mathbf{F}_{\mathrm{\!BB2}}^{\mathrm{opt}} \right\|}_{F}} \\ 
 & \quad \quad \quad \quad \quad \quad \,\le {{\left\| {\mathbf{E}_{\mathrm{1}}}\!-\!\mathbf{F}_{\mathrm{\!RF2}}^{\mathrm{opt}}\mathbf{F}_{\mathrm{\!BB2}}^{\mathrm{opt}} \right\|}_{\!F}}\!+\!\left| 1\!-\!\frac{1}{\gamma } \right|{{\left\| \mathbf{F}_{\mathrm{\!RF2}}^{\mathrm{opt}}\mathbf{F}_{\mathrm{\!BB2}}^{\mathrm{opt}} \right\|}_{F}} \\ 
 & \quad \quad \quad \quad \quad \quad \,\le \delta +\left| \gamma -1 \right|\sqrt{\frac{{{P}_{1}}}{{{P}_{2}}}}{{\left\| {\mathbf{E}_{\mathrm{1}}} \right\|}_{F}} \\ 
 & \quad \quad \quad \quad \quad \quad \,\le \left( 1+\left| \frac{\gamma -1}{\gamma -\sqrt{{{{P}_{1}}}/{{{P}_{2}}}\;}} \right| \right)\delta \ .  \\ 
\end{aligned}
\label{eq52}
\end{equation}
\end{small}Due to {\small $\mathbf{F}_{\mathrm{RF2}}^{\mathrm{opt}}\mathbf{F}_{\mathrm{BB2}}^{\mathrm{opt}}\approx {\mathbf{E}_{\mathrm{1}}}$}, we have
\begin{small}
\begin{equation}
  \sqrt{\frac{{{P}_{1}}}{{{P}_{2}}}}\!=\!\frac{\sqrt{{{N}_{\mathrm{s}}}}}{{{\left\| \mathbf{F}_{\mathrm{\!RF}}^{\mathrm{opt}}\mathbf{F}_{\mathrm{\!BB}}^{\mathrm{opt}} \right\|}_{F}}}\frac{{{\left\| \mathbf{F}_{\mathrm{\!RF2}}^{\mathrm{opt}}\mathbf{F}_{\mathrm{\!BB2}}^{\mathrm{opt}} \right\|}_{F}}}{{{\left\| {\mathbf{E}_{\mathrm{1}}} \right\|}_{F}}}\!\approx \!\frac{\sqrt{{{N}_{\mathrm{s}}}}}{{{\left\| \mathbf{F}_{\mathrm{\!RF}}^{\mathrm{opt}}\mathbf{F}_{\mathrm{\!BB}}^{\mathrm{opt}} \right\|}_{F}}}\!=\!\frac{1}{\gamma }\ .
  \label{eq53}
\end{equation}
\end{small}Substituting (\ref{eq53}) into (\ref{eq52}), the conclusion is obtained:
\begin{small}
\begin{equation}
  {{\left\| {\mathbf{E}_{\mathrm{1}}}-\mathbf{F}_{\mathrm{RF2}}^{\mathrm{opt}}\mathbf{\hat{F}}_{\mathrm{BB2}}^{\mathrm{opt}} \right\|}_{F}}\le \left( 1+\left| \frac{\gamma }{\gamma +1} \right| \right)\delta <2\delta \ .
  \label{eq54}
\end{equation}
\end{small}

Similarly, it is evident that if { \small ${{\left\| {{\mathbf{E}}_{\mathrm{2}}}-\mathbf{F}_{\mathrm{RF1}}^{\mathrm{opt}}\mathbf{F}_{\mathrm{BB1}}^{\mathrm{opt}} \right\|}_{F}}\le \delta $} holds, {\small ${{\left\| {{\mathbf{E}}_{\mathrm{2}}}-\mathbf{F}_{\mathrm{RF1}}^{\mathrm{opt}}\mathbf{\hat{F}}_{\mathrm{BB1}}^{\mathrm{opt}} \right\|}_{F}}\le 2\delta $} will also hold: replacing parameters {\small $\sqrt{{{P}_{1}}}$}, {\small $\sqrt{{{P}_{2}}}$}, {\small$\mathbf{F}_{\mathrm{BB2}}^{\mathrm{opt}}$}, {\small $\mathbf{F}_{\mathrm{RF2}}^{\mathrm{opt}}$} and {\small ${\mathbf{E}_{\mathrm{1}}}$} in (\ref{eq50}) to (\ref{eq54}) with parameters {\small $\sqrt{{{N}_{\mathrm{s}}}}$}, {\small $\sqrt{{{N}_{\mathrm{s}}}}$}, {\small $\mathbf{F}_{\mathrm{BB1}}^{\mathrm{opt}}$}, {\small $\mathbf{F}_{\mathrm{RF1}}^{\mathrm{opt}}$} and {\small ${{\mathbf{E}}_{\mathrm{2}}}$} respectively, (\ref{eq53}) will transform into {\small $\sqrt{{{{P}_{1}}}/{{{P}_{2}}}\;}=1$}. Substitute it into (\ref{eq52}) and then we have {\small ${{\left\| {{\mathbf{E}}_{\mathrm{2}}}-\mathbf{F}_{\mathrm{RF1}}^{\mathrm{opt}}\mathbf{\hat{F}}_{\mathrm{BB1}}^{\mathrm{opt}} \right\|}_{F}}\le \left( 1+\left| \frac{\gamma -1}{\gamma -1} \right| \right)\delta =2\delta $}, resulting in a similar conclusion in \cite{yu2016alternating} which is a special case in (\ref{eq52}) under conditions of {\small $\sqrt{{{P}_{1}}}=\sqrt{{{P}_{2}}}=\sqrt{{{N}_{\mathrm{s}}}}$}. 
}\hfill$\blacksquare$

\linespread{0.938} 
\bibliographystyle{IEEEtran}
\bibliography{references}

\end{document}